\documentclass[fleqn,usenatbib]{mnras}

\usepackage{newtxtext,newtxmath}

\usepackage[T1]{fontenc}
\usepackage{fontawesome}

\DeclareRobustCommand{\VAN}[3]{#2}
\let\VANthebibliography\thebibliography
\def\thebibliography{\DeclareRobustCommand{\VAN}[3]{##3}\VANthebibliography}

\usepackage{graphicx}	
\usepackage{amsmath}	
\usepackage{CJKutf8}
\usepackage{hyperref}
\newcommand{\orcid}[2]{\href{http://orcid.org/#2}{#1} \includegraphics[width=12pt]{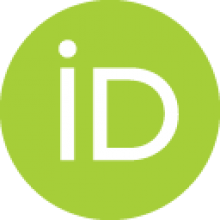}}

\title[Measuring the X-ray luminosities of DESI groups]{Measuring the X-ray luminosities of DESI groups from eROSITA Final Equatorial-Depth Survey: I. X-ray luminosity - halo mass scaling relation}

\author[Zheng et al.]{
\orcid{Yun-Liang Zheng \begin{CJK}{UTF8}{gkai}(郑云亮)\end{CJK}}{0000-0002-5632-9345},$^{1}$\thanks{E-mail: zhengyunliang@sjtu.edu.cn}
\orcid{Xiaohu Yang \begin{CJK}{UTF8}{gkai}(杨小虎)\end{CJK}}{0000-0003-3997-4606},$^{1,2}$
\orcid{Min He \begin{CJK}{UTF8}{gkai}(何敏)\end{CJK}}{0000-0001-6139-7660},$^{3}$
\orcid{Shi-Yin Shen \begin{CJK}{UTF8}{gkai}(沈世银)\end{CJK}}{0000-0002-3073-5871},$^{4,5}$\newauthor
\orcid{Qingyang Li \begin{CJK}{UTF8}{gkai}(李清洋)\end{CJK}}{0000-0003-0771-1350}$^{1}$ 
and Xuejie Li \begin{CJK}{UTF8}{bkai}(李雪劼)\end{CJK}$^{1}$
\\
\\
$^{1}$Department of Astronomy, School of Physics and Astronomy, and Shanghai Key Laboratory for Particle Physics and Cosmology, \\Shanghai Jiao Tong University, Shanghai 200240, People's Republic of China\\
$^{2}$Tsung-Dao Lee Institute, and Key Laboratory for Particle Physics, Astrophysics and Cosmology, Ministry of Education, \\Shanghai Jiao Tong University, Shanghai 200240, People's Republic of China\\
$^{3}$CAS Key Laboratory of Optical Astronomy, National Astronomical Observatories, Chinese Academy of Sciences, Beijing 100101, China\\
$^{4}$Key Laboratory for Research in Galaxies and Cosmology, Shanghai Astronomical Observatory, \\Chinese Academy of Sciences, 80 Nandan Road, Shanghai, 200030, People's Republic of China\\
$^{5}$Key Lab for Astrophysics, Shanghai, 200034, People's Republic of China
}

\date{Accepted XXX. Received YYY; in original form ZZZ}

\pubyear{2023}

\begin{document}

\label{firstpage}
\pagerange{\pageref{firstpage}--\pageref{lastpage}}
\maketitle

\begin{abstract}
We use the eROSITA Final Equatorial-Depth Survey (eFEDS) to measure the rest-frame 0.1-2.4 keV band X-ray luminosities of $\sim$ 600,000 DESI groups using two different algorithms in the overlap region of the two observations. These groups span a large redshift range of $0.0 \le z_g \le 1.0$ and group mass range of $10^{10.76}h^{-1}M_{\odot} \le M_h \le 10^{15.0}h^{-1}M_{\odot}$. (1) Using the blind detection pipeline of eFEDS, we find that 10932 X-ray emission peaks can be cross matched with our groups, $\sim 38 \%$ of which have signal-to-noise ratio $\rm{S}/\rm{N} \geq 3$ in X-ray detection. Comparing to the numbers reported in previous studies, this matched sample size is a factor of $\sim 6$ larger. (2) By stacking X-ray maps around groups with similar masses and redshifts, we measure the average X-ray luminosity of groups as a function of halo mass in five redshift bins.  We find, in a wide halo mass range, the X-ray luminosity, $L_{\rm X}$, is roughly linearly proportional to  $M_{h}$, and is quite independent to the redshift of the groups.  (3) We use a Poisson distribution to model the X-ray luminosities obtained using two different algorithms and obtain best-fit $L_{\rm X}=10^{28.46\pm0.03}M_{h}^{1.024\pm0.002}$ and $L_{\rm X}=10^{26.73 \pm 0.04}M_{h}^{1.140 \pm 0.003}$ scaling relations, respectively. The best-fit slopes are flatter than the results previously obtained, but closer to a self-similar prediction.
\end{abstract}




\section{Introduction}
A galaxy group\footnote{In this paper, we refer to a system of galaxies as a group regardless of its mass and richness (i.e., rich clusters or groups with a single galaxy member)} is a concentration of galaxies assumed to be embedded within an extended dark matter halo, providing cosmological probes of the spatial distribution and growth history of large-scale structure. The relatively high density make galaxy group an ideal site for studying the formation and evolution of galaxies within the framework of hierarchical paradigm. However, from observational point of view, the membership of group systems are not easy to determine because dark matter halos cannot be observed directly. Therefore, numerous group finding algorithms have been developed to identify the galaxy groups either from photometric or spectroscopic surveys: e.g., \citet{Yang..2005} and \citet{Einasto..2007} from the 2-degree Field Galaxy Redshift Survey; \citet{Weinmann..2006}, \citet{Yang..2007, Yang..2012}, \citet{Tempel..2014, Tempel..2017}, \citet{Munoz-Cuartas.Muller2012}, and \citet{Rodriguez.Merchan2020} from the Sloan Digital Sky Survey; \citet{Lu..2016} and \citet{Lim..2017} from the the Two Micron All Sky Survey; \citet{Robotham..2011} from the Galaxy and Mass Assembly Survey; \citet{WangKai..2020} from the zCOSMOS Survey; \citet[][hereafter Y21]{Yang..2021} from the DESI Legacy Image Surveys (LS). These group catalogs provide group systems that have relatively reliable membership determination, which is important for studying the galaxy evolution driven by the environment \citep[e.g.,][]{PengYJ..2010, PengYJ..2012, Wetzel..2012, WangEnci..2018, LiuCX..2019, Davies..2019}. Galaxy interactions such as mergers and close encounters are crucial mechanisms for the transformation of galaxy population in group environment \citep[e.g.,][]{Ellison..2008, Ellison..2013, Patton..2016, Pearson..2019, Feng..2019, Feng..2020}. 

Besides the galaxy members, another known baryonic component retained within the group systems is the intragroup medium (IGM), which is a diffuse gas hot enough to emit X-rays mainly through bremsstrahlung. This hot IGM  interacts with the gas within the infalling galaxies leading to the removal of the cold gas that fuels the star formation activity. This IGM can also alter the properties of galaxies which are already in the groups. The density, temperature and entropy profiles of the IGM might decode the entire thermal history of that group. Albeit with these advantages, blind X-ray detection of groups typically has a low efficiency. Unlike a typical massive galaxy group having X-ray emission extend up to several Mpc, less massive groups often show lower and flatter X-ray surface brightness \citep[e.g.,][]{Mulchaey..2000, Santos..2008, Rasia..2013, Lovisari..2017, Yuan.Han2020}. In order to pursue the gas properties in lower mass groups, large area and deep X-ray observations are always in great demand. Apart from this, an alternative way to enhance the detection limit is to use  prior information about the position and size of each group that can be obtained from e.g., optical observations. 

Since the first X-ray all-sky survey performed with the ROSAT telescope, X-ray properties of the optical-selected groups have been extensively studied \citep[e.g.,][]{Donahue..2001, Mulchaey..2003, Brough..2006, DaiXY..2007, Popesso..2007, Shen..2008, WangLei..2014, Paper3}. Although a number of subsequent X-ray surveys reaching fluxes about three dex fainter than RASS have been achieved, the sky coverage of most of them is smaller than $\sim 100$ deg$^2$, only a small number of the group systems have been analysed \citep[][]{Rasmussen..2006, Andreon.Moretti2011, Hicks..2013, Pearson..2017}. 
Based on these X-ray observations, a number of X-ray luminosity v.s. halo mass relations were obtained. However, these relations are not yet converged especially at the low mass end due to the insufficient observations \citep[][and references therein]{Lovisari..2021}.  

Recently, eROSITA offers the next major step forward for studying the X-ray properties in $0.2 - 10$ keV for group systems that can be identified based on the galaxy surveys with large sky coverages. eROSITA will finish the entire sky scanning for eight times, using an array of seven aligned telescope modules (TMs). Before the complete all-sky survey, the eROSITA Final Equatorial-Depth Survey (eFEDS) was designed to test the capability of eROSITA. This field overlaids on a various of deep optical/NIR surveys such as HSC Wide Area Survey \citep{Aihara..2018}, KIDS-VIKINGS \citep{Kuijken..2019}, DESI Legacy Imaging Survey \citep{Dey..2019}, and so on. In addition, eFEDS also overlaps the region of XMM-ATLAS survey \citep{Ranalli..2015} which can provide a useful dataset for comparison and test the reliability of the results obtained by eFEDS. This data set, combined with the group catalogs recently constructed by Y21 from the DESI LS observations within redshift range $0.0<z<1.0$, will thus provide us an unique opportunity to measure the X-ray luminosity around galaxy groups that span both large redshift and halo mass ranges. It will enable us to better constrain the X-ray luminosity v.s. halo mass scaling relation in a much  larger redshift and halo mass range.

This paper is organized as follows. In section~\ref{sec:data}, we describe the data used in this work. In section~\ref{sec:xray}, we perform the X-ray luminosity measurement for the DESI groups, and test the reliability of our X-ray luminosity measurement by comparing to existing X-ray group catalogs. We investigate the scaling relation between the X-ray luminosity and group mass in section~\ref{sec:general}. Finally, we draw our conclusions in section~\ref{sec:conclusion}. 
Throughout this paper, we assume a flat $\Lambda$ cold matter cosmology with parameters: $\Omega_{\rm m} = 0.315$ and $H_0 = 100h$ km s$^{-1}$Mpc$^{-1}$ with $h=0.7$. If not specified otherwise, the X-ray luminosity $L_{\rm X}$ and fluxes $f_{\rm X}$ are given in the range of $0.1-2.4$ keV band.

\begin{figure*}
\centering
    \includegraphics[width=.95\hsize]{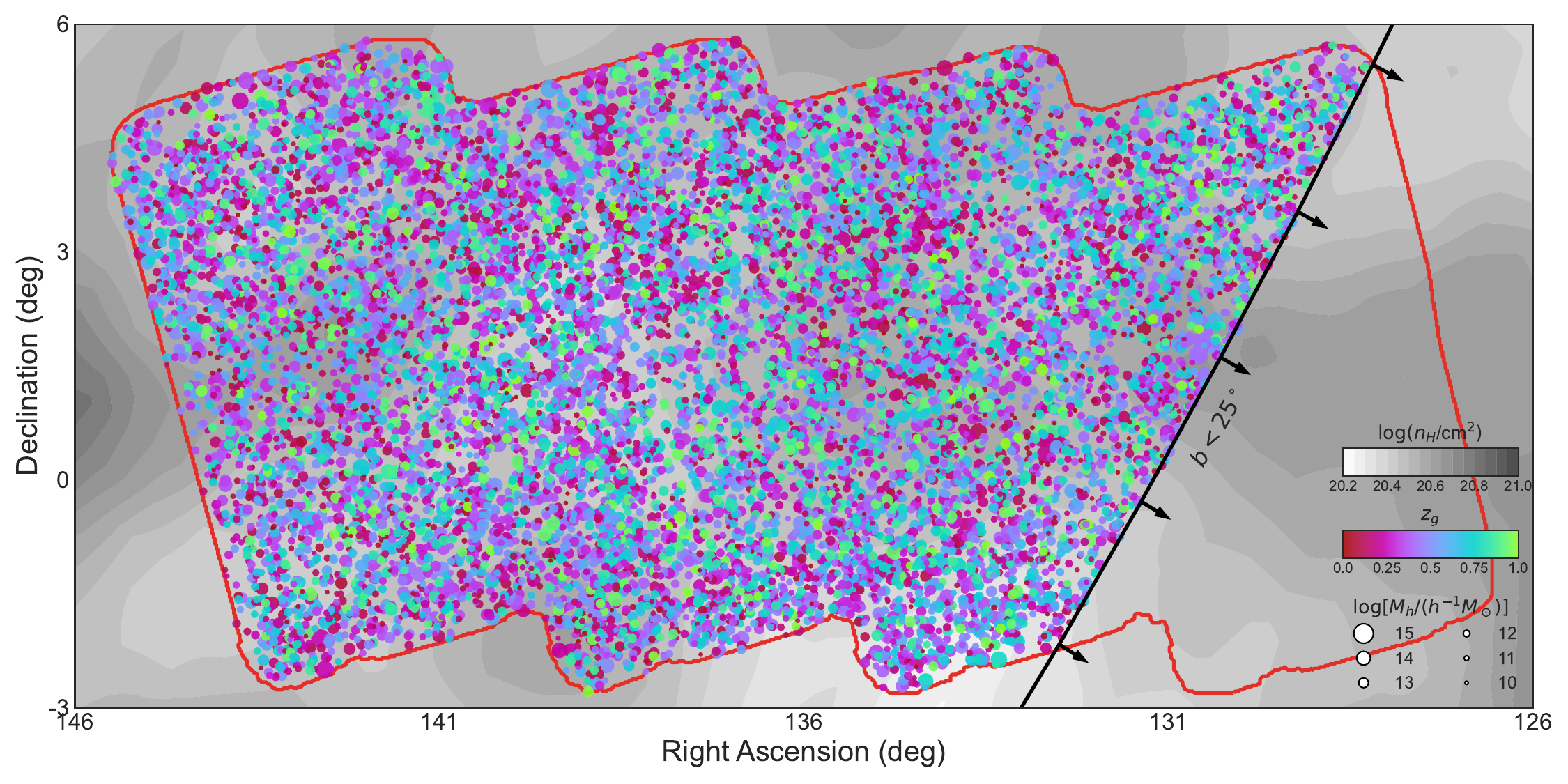}
  \caption{The distribution of the DESI groups overlaid on the footprint of eFEDS (red lines enclosed). The dots represent the DESI groups with blind-detected X-ray centers color-coded by their $z_g$. The radius of the dots corresponds to the $M_h$ of each galaxy group. The contour show the galactic hydrogen column density ,$N_{\rm H}$, along the line of sight to each point given by HEALPIX resampling of Leiden/Argentine/Bonn Survey of Galactic HI \citep[][]{Kalberla..2005}.}
  \label{fig:eFEDS}
\end{figure*}

\section{The DESI Group Catalog} \label{sec:data}

The group catalog used in this work is taken from Y21, which extended the halo-based group finder developed by \citet{Yang..2005} and applied it to the DESI Legacy Imaging Surveys. Every galaxy within the photometric redshift range of $0.0 < z < 1.0$ and $z-$band magnitude brighter than $m_{z} = 21$ mag was assigned to a unique group. This catalog has been removed the area within $|b| \leq 25^\circ$ to avoid the regions of higher stellar density. The redshift of each galaxy is taken from the random-forest-algorithm-based photometric redshift estimation from the \textit{Photometric Redshifts for the Legacy Surveys} \citep[PRLS,][]{ZhouRP..2021}, with a typical redshift error of $\sigma_z/(1+z) \sim 0.02$. To ensure the redshift information as accurate as possible, a small fraction of the redshifts have been replaced by the avaliable spectroscopic redshifts to date \citep[see more detail in][]{Yang..2021}. This group catalog has been further updated based on the galaxy catalog which has been updated to DR9, containing $\sim 100$ million groups with $\sim 120$ million galaxy members having five-band photometries ($g$, $r$, $z$, $W1$, $W2$), with a sky coverage of $\sim 18200$ deg$^{2}$. 

The sky coverage of eFEDS is $\sim 140$ deg$^2$, only $\sim 100$ deg$^2$ of which has overlaid on the footprint of DESI galaxies as shown in Figure~\ref{fig:eFEDS}. In total, there are $\sim 600,000$ DESI groups overlapped with eFEDS and can be used to perform the X-ray luminosity measurements. In Y21, the group dark matter halos are defined as having an overdensity of 180 times the background density of the universe. The halo mass ($M_h$) of each group has been estimated based on abundance matching between the total group luminosity and halo mass assuming a Planck18 cosmology \citep{Planck18}. The $M_h$ has an uncertainty of $\sim 0.2$ dex at high mass end ($M_h \gtrsim 10^{14}h^{-1}M_{\odot}$), increasing to $\sim 0.4$ dex at $M_h \sim 10^{12.3}h^{-1}M_{\odot}$ and then decreasing to $\sim 0.3$ dex at $M_h \sim 10^{11}h^{-1}M_{\odot}$. The angular virial radius, $\theta_{180}$, is calculated using
\begin{equation}
\theta_{180} = {\left(\cfrac{M_{h}}{\frac{4\pi}{3} \cdot 180 \Omega_{\rm m} \cdot \frac{3 H_0^2}{8 \pi G}}\right)}^{1/3} \cdot D_{\rm c}^{-1}, 
\end{equation}
where $D_{\rm c}$ is the comoving distance of that group.

\begin{figure*}
\centering
    \includegraphics[width=1.\hsize]{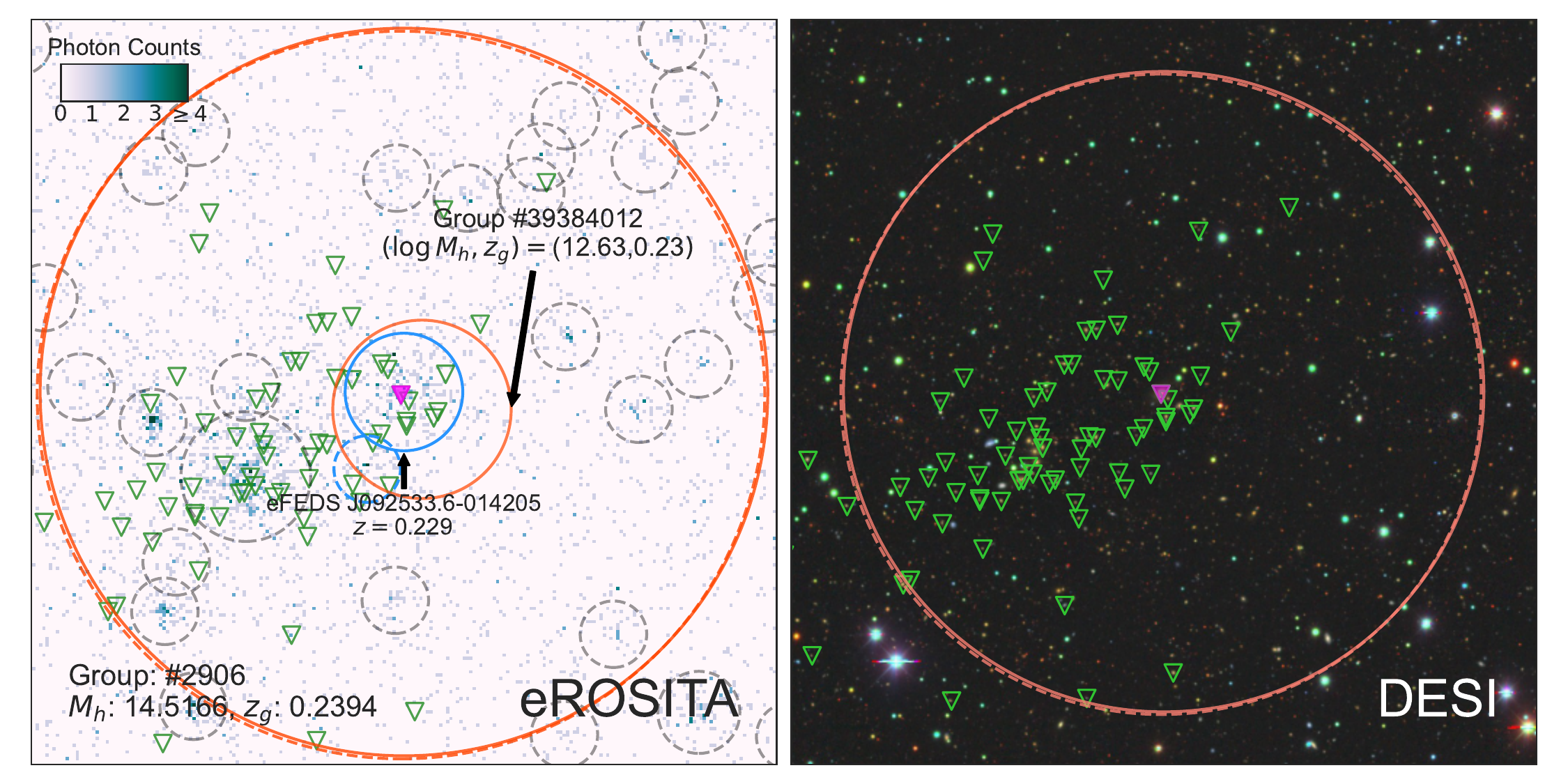}
  \caption{An example eFEDS image (left panel) for a blind-detected source (blue solid circle) overplotted with the probable DESI groups that might host it. The red solid circles show the regions within a distance of $R_{180}$ from the BGG of each candidate group. In this example, we regard the most massive one (group \#2906) hosts that X-ray emission. The neighbor X-ray emission (blue dashed circles) is also considered as part of the extended X-ray emission from the same group, while the others (grey dashed circles) are regarded as contaminants. The red dashed circle represents the region within a distance of $R_{180}$ but re-centered on that X-ray emission. The right panel shows the corresponding DESI image. In both panels, the BGG and satellites of group \#2906 are marked by magenta filled and green opened circles, respectively. }
  \label{fig:example}
\end{figure*}

\section{X-ray Detection} \label{sec:xray}
We use the public eROSITA data from the eFEDS field. The eFEDS field is divided into four sections, each of which has a separate event list. In this section, we reduced the data with eROSITA Science Analysis Software System (\texttt{eSASS}).

Following \citet{Brunner..2022}, we apply the astrometric corrections to the observation attitude and then recalculate the event coordinates using the \texttt{eSASS} tasks \texttt{evatt} and \texttt{radec2xy}. Next, we convert the event list into an image with \texttt{evtool} command and generate the corresponding exposure map with \texttt{expmap} command. In this work, the imaging analysis is performed in the soft X-ray band ($0.2-2.3$ keV) with average exposure time of $\sim 1.2$ ks after correcting for telescope vignetting across most of the field\footnote{As pointed by \citet{Brunner..2022}, some events could not be used due to an unrecognized malfunction of the camera electronics, resulting in a reduced exposure depth in the affected areas (see figure 1 in \citealt{Brunner..2022}). Such events do not exist in the calibrated event files.}. 

Because of the group position and halo radius have already been determined, we use these information when analysing their X-ray properties. The algorithm we use to measure the X-ray of each DESI group is similar to those of \citet[][hereafter S08]{Shen..2008}, \citet[][hereafter W14]{WangLei..2014}, and \citet{Paper3} but with a set of improvement. In section~\ref{sec:erosita-sources}, we determine the X-ray center for each DESI group. In section~\ref{sec:stack}, we perform the stacks for groups without blind-detected centers at different $M_h$ and $z_g$ bins. In section~\ref{sec:cts} and \ref{sec:xlums}, we obtain the source count rate and X-ray luminosity for each DESI group using different algorithms. In section~\ref{sec:comparison}, we compare our results with the previous studies.

\subsection{Determine the X-ray center for each DESI group} \label{sec:erosita-sources}
Based on the eFEDS maps, one can detect the possible X-ray peaks that are emitted from various kinds of X-ray sources. \citet{Brunner..2022} present a primary catalog of 27910 X-ray sources detected in the $0.2-2.3$ keV band with detection likelihood $\mathscr{L}_{\rm det} \ge 6$ and a supplementary catalog of 4774 X-ray sources detected in the same band but with detection likelihood of $5 \le \mathscr{L}_{\rm det} < 6$. Almost all of the blind-detected targets are point-like sources with extent likelihood $\mathscr{L}_{\rm ext} = 0$, while only 542 sources with extent likelihood $\mathscr{L}_{\rm det} \ge 6$ are treated as X-ray emission from massive galaxy groups \citep{LiuA..2022}. However, \citet{Bulbul..2022} pointed that high redshift galaxy clusters or nearby groups hosting bright AGNs might be potentially misclassified as point sources by the pipeline due to the sizeable point-spread function of eROSITA, and select a catalog of 346 X-ray sources with extent likelihood $\mathscr{L}_{\rm ext} = 0$ that are indeed galaxy groups with mass of $10^{13} - 4.5 \times 10^{14} M_{\odot}$ in disguise from the primary catalog. Both studies apply a multi-component matched filter (MCMF) cluster confirmation tool \citep{Klein..2018, Klein..2019} to determine the redshifts of 888 X-ray clusters in total. This implies that more faint X-ray point sources might be emitted from smaller or more distant galaxy groups. 

In order to remove the signals from contaminants, we need to mask out all of the background or foreground sources when calculating the count rates for each group. \citet{Salvato..2022} have presented the identification of the counterparts to the $\mathscr{L}_{\rm ext} = 0$ sources in primary catalog and classify them into `secure galactic', `likely galactic', `secure extragalactic', and `likely extragalactic' sources. Most of the point-like sources belong to the last two cases, we thus regard all of the targets in supplementary catalog as extragalactic sources. Then we cross-match the extragalactic sources to SDSS DR16 quasar catalog \citep{sdss16q} within a tolerance of 10 arcsec and identify $\sim 2300$ background quasars. Besides the 888 X-ray groups identified by \citet{LiuA..2022} and \citet{Bulbul..2022}, the remaining galactic sources and quasars are contaminants that are not associated with any group systems. 

For the remaining extragalactic X-ray sources, we match each of them to DESI groups within a maximum separation of $0.3 R_{180}$ and $|z-z_{\rm MCMF}| \le 0.05$ (if it has redshift assigned by \citet{LiuA..2022} and \citet{Bulbul..2022}) from the BGG of each DESI group. Owing to the fact that most of the extragalactic X-ray sources have numerous DESI groups matched, we regard the most massive one as the host of that X-ray emission for simplicity. If no groups matched, this X-ray source might be emitted from the targets with $z \gtrsim 1$. For the groups with numerous X-ray sources matched, we regard the one closest to the BGG as X-ray center of that group, the matched sources within $0.3 R_{180}$ from the X-ray center as parts of the extended X-ray emission, and the others beyond $0.3 R_{180}$ would be re-matched to the second most massive group satisfying the aforementioned criteria. This iterative process goes on until there is no further change in the group matching. In the left panel of Figure~\ref{fig:example}, we show an example image for the DESI groups matched by an X-ray source (blue dashed circle), there are two groups might host that X-ray emitter. According to our criteria, this emitter is more likely to be the X-ray center for the more massive candidate. Finally, $10932$ DESI groups host at least one blind-detected sources. For most of the other DESI groups, we assign the position of BGGs as their X-ray centers. In the next section, we will check whether the position of the BGG is close to a peak in X-ray emission.

\begin{figure*}
\centering
    \includegraphics[height=.47\vsize]{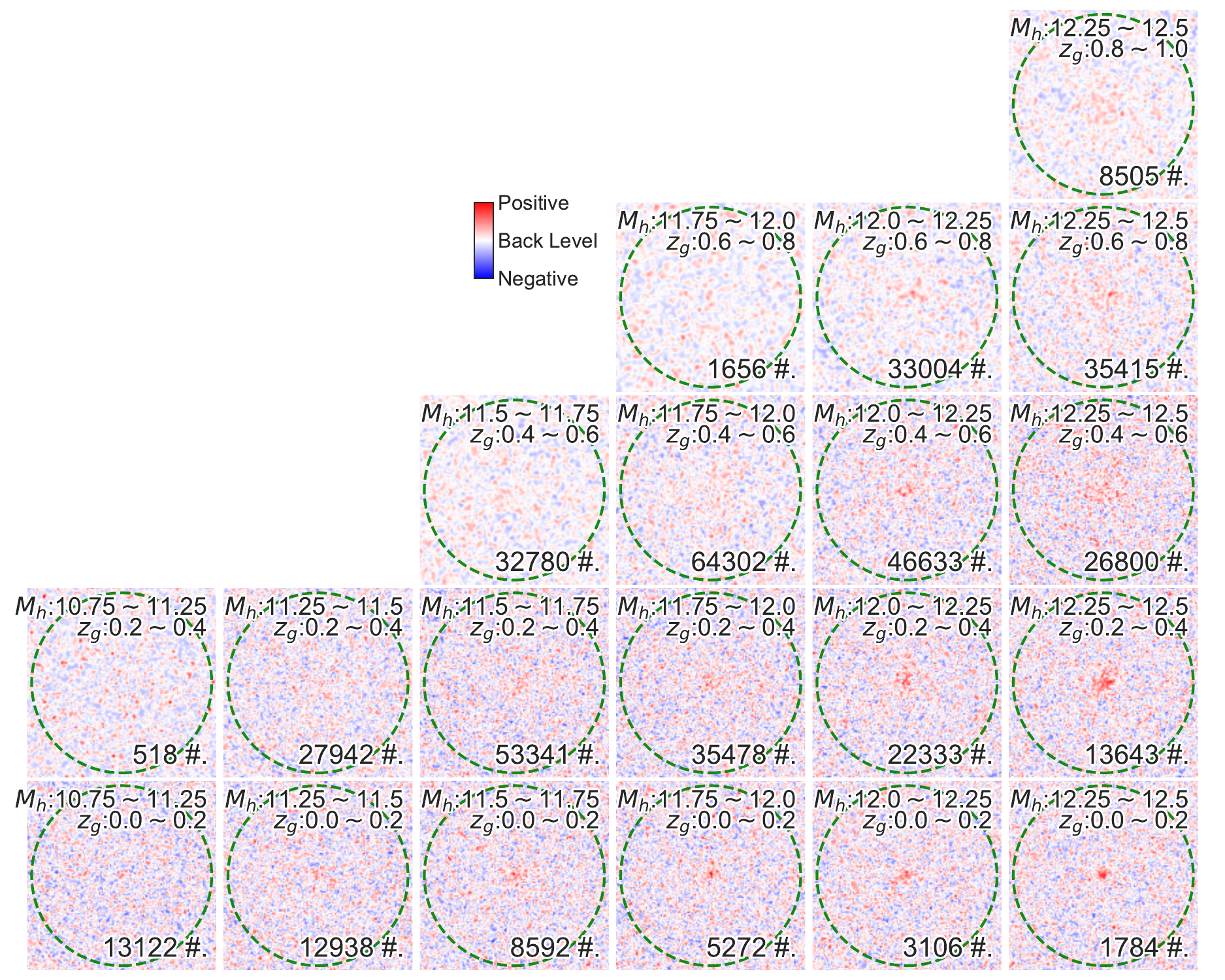}
    \includegraphics[height=.47\vsize]{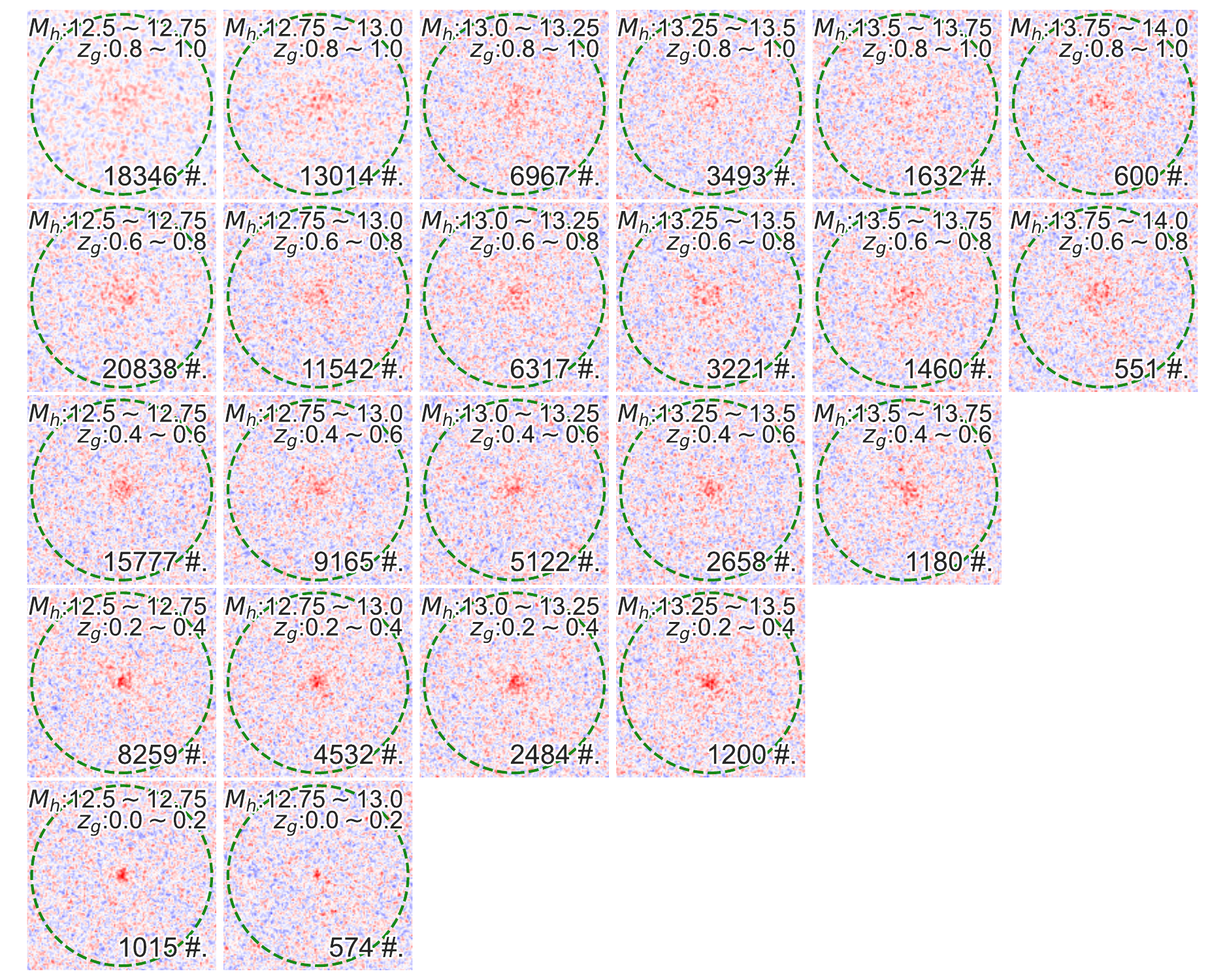}
  \caption{Stacked eFEDS images of DESI groups without resolved X-ray centers in different $M_h$ and $z_g$ bins. The dashed circles represent the regions within a radius of $R_{180}$. Only the data bins with at least 500 groups are shown here. }
  \label{fig:stack}
\end{figure*}

\begin{figure*}
\centering
    \includegraphics[height=0.46\vsize]{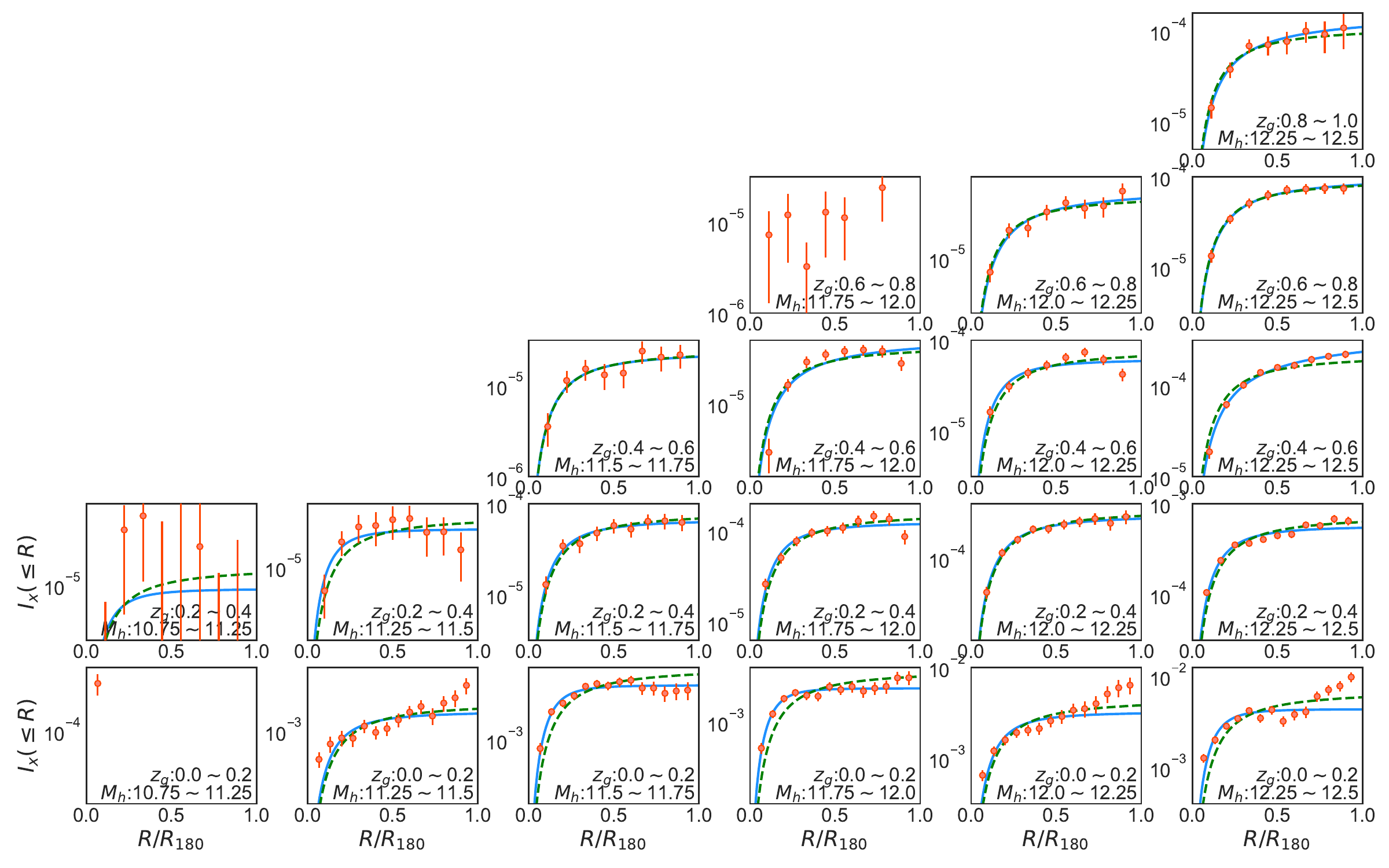}
    \includegraphics[height=0.46\vsize]{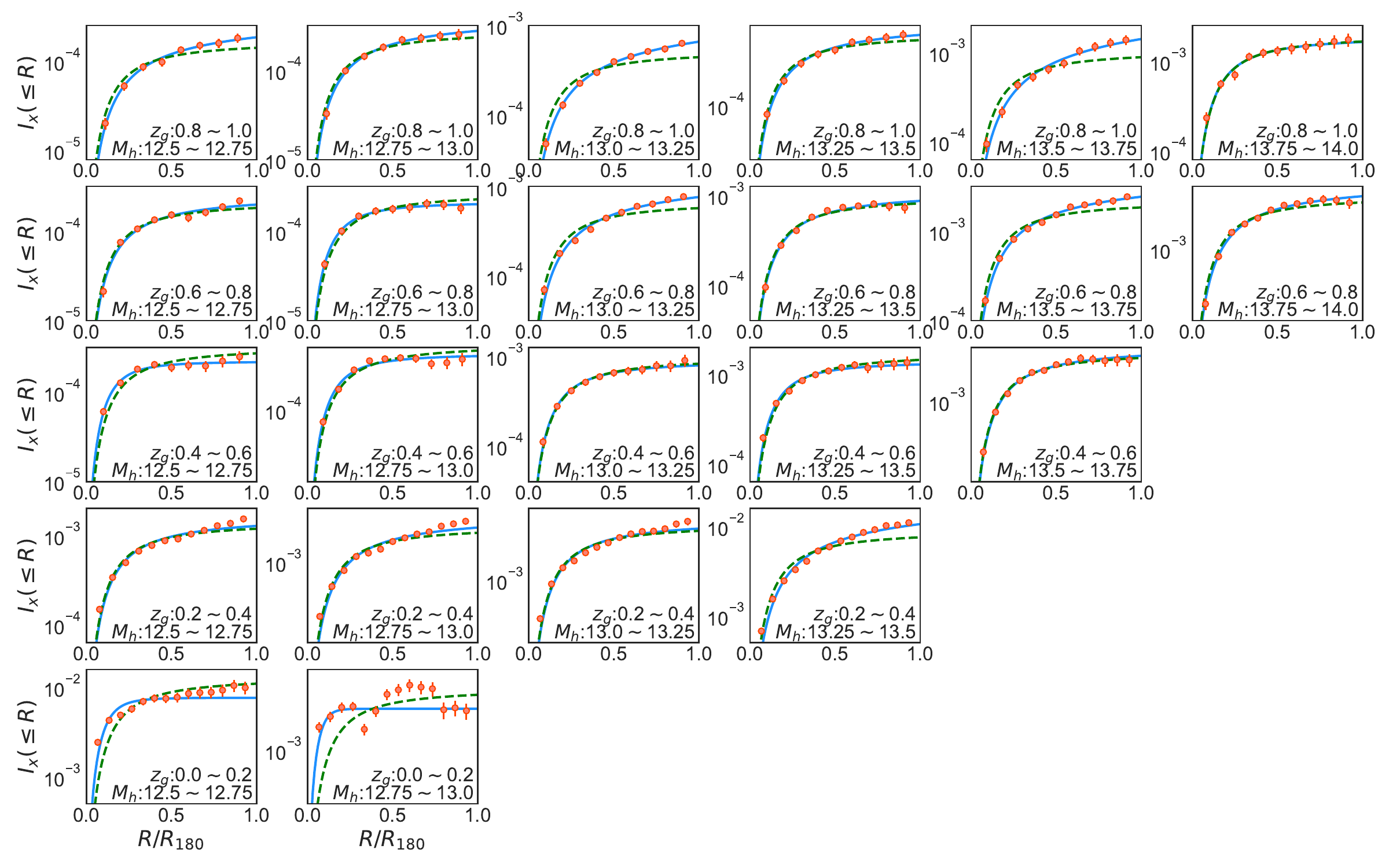}
  \caption{Stacked cumulative surface brightness profiles as a function of $R/R_{180}$ of DESI groups without resolved X=ray centers in different $M_h$ and $z_g$ bins (red points with error bars). The blue solid lines are the best-fit cumulative $\beta-$model for each stack, while the green dashed lines represent the same fitting but with a fixed $\beta = 2/3$ for reference. Only the data bins with at least 500 groups are shown here.}
  \label{fig:profile}
\end{figure*}

\subsection{X-ray Stacks} \label{sec:stack}
Before individual measurement, we perform the stacks for DESI groups without resolved X-ray emission first. As discussed in the last section, the net photon count is very low for most of the DESI groups on eFEDS map. To show the reliability of the determination for X-ray center, we produce the stacked images for the groups by rescaling the data for each group to a common size. We donot weight the photon by the square of ratio between the group luminosity distance and an arbitrary fixed value, because the redshift bin is relatively small for each subsample. We binned the X-ray images with pixel size of 4" and mask out all of the contaminants.

In figure~\ref{fig:stack}, we show the stacked images of DESI groups without blind-detected X-ray center at different $M_h$ and $z_g$. Note that we only show the data bins with at least 500 groups. There is no doubt that the signals are clearer than individual measurements. We see the X-ray excess, although not very significant, around the X-ray center for most of the stacks, implying that BGG can well represent the X-ray peak of a group system. The central excess appears clearer with increasing $M_h$ at a given redshift bin, because the X-ray emission is much more evident in massive systems, while such excess appears more fuzzy for distant groups due to the flux limit and resolution.

For each stack, we compute the background using an annuli at $1.5 R_{180} < R < 2.0 R_{180}$ and derive the corresponding surface brightness profile. The surface brightness profile for X-ray emission of group system can be well described by an empirical $\beta-$profile \citep{beta-model}:
\begin{equation}
\mu(R/R_{180}) = \mu_0{\left[1+{\left(\frac{R}{R_c}\right)}^2\right]}^{-3\beta+0.5},
\end{equation}
where $\mu_0$ is the central surface brightness and $R_c = 0.18R_{180}$ is the core radius. 

The parameters ($\mu_0$, $\beta$) for each stack are not fitted using the above form directly because the observed data of the innermost bins almost determines the fitting results. We choose the better determination of these parameters using the cumulative form of $\beta-$model. The cumulative source count rate as a function of radius is computed by integrating the net source counts in concentric ring. In figure~\ref{fig:profile}, we show the cumulative source count rate and the best-fitting $\beta-$profile for DESI groups without blind-detected X-ray center at different $M_h$ and $z_g$. As can be seen, the surface brightness distribution can be well characterized by $\beta-$profile except some least massive bins. We also plot the results with a fixed value of $\beta = 2/3$, which have been extensively adopted \citep[e.g.,][]{Arnaud..1999, Reiprich..2002, Ettori..2004, Maughan..2006, Hicks..2008}, for reference. In most of the stacks, the best-fit results show slightly less concentrated ($0.4 \lesssim \beta < 2/3$) compared with the references, partly due to the off-center effect that the X-ray peak is not necessarily coincident with the position of BGG. In this work, we focus on the X-ray flux of each group, although the slight off-center effect might lower the value of $\beta$, the overall count rate will not be affected much. Moreover, the fluctuations in background estimate might lower the signal-to-noise (S/N) of cumulative count rate in the outermost bins and enlarge the error of $\beta$. For security, we only calculate the count rates enclosed within $R_{\rm X} = 0.5R_{180}$ and make a $\beta-$profile extension correction to make up the X-ray luminosity missed in the range $R_{\rm X} \le R \le R_{180}$ \citep[see][]{Bohringer..2000, Shen..2008, WangLei..2014} for each individual group. The extension correction factor is not very sensitive to the value of $\beta$ when $\beta \gtrsim 0.4$ ($\lesssim 0.3$ dex), and we adopt a fixed value of $\beta = 2/3$.

\begin{figure}
    \centering
    \includegraphics[width=1.\hsize]{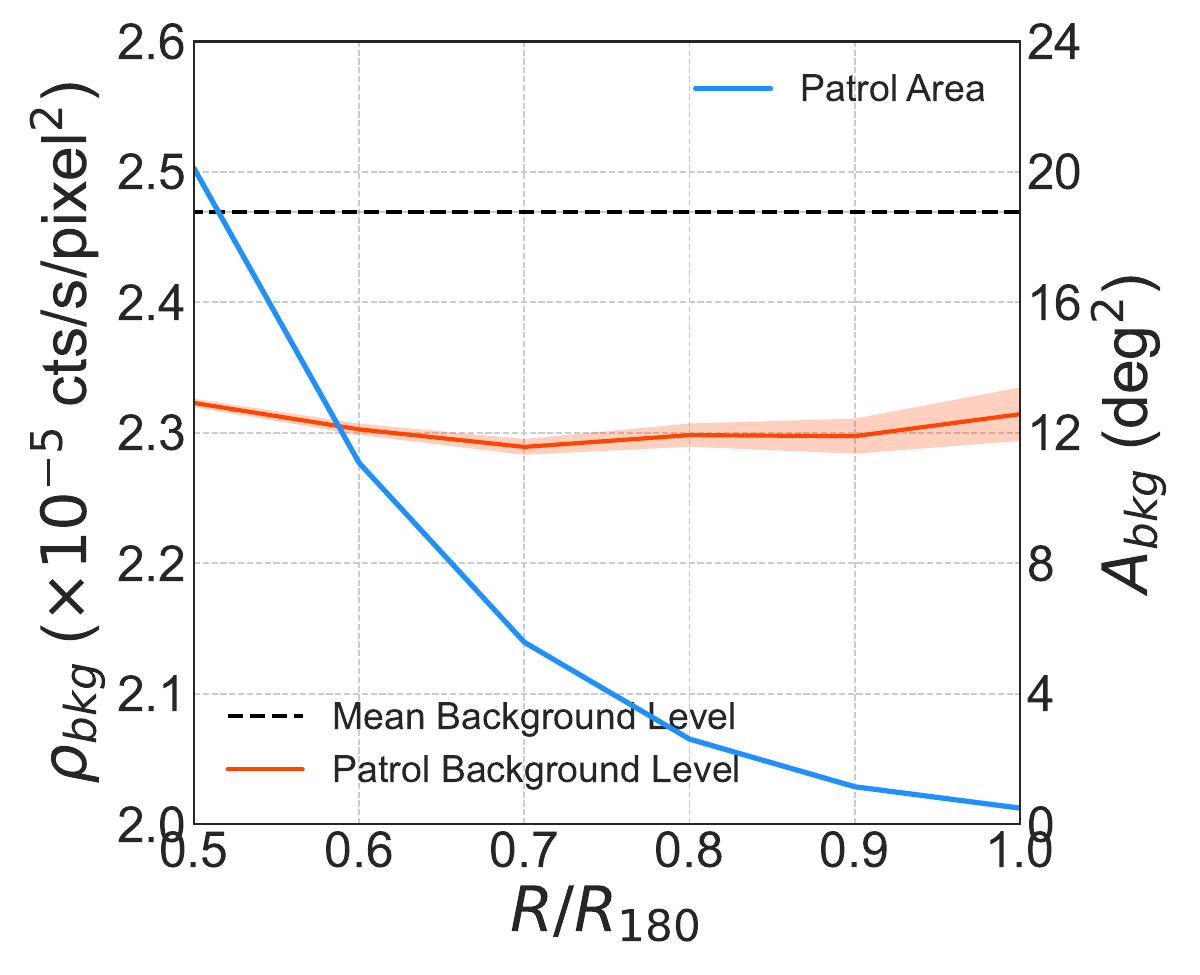}
    \caption{The red solid line with shaded region represents the background count rate density with errors for the patrol area based on different value of aperture radius (from 0.5 to 1.0 $R_{180}$, see details in section~\ref{sec:cts}), the corresponding patrol area is shown in blue solid line. For reference, we also plot the mean background count rate density in black dashed line.}
    \label{fig:patrol}
\end{figure}

\begin{figure*}
\centering
  \includegraphics[width=.95\hsize]{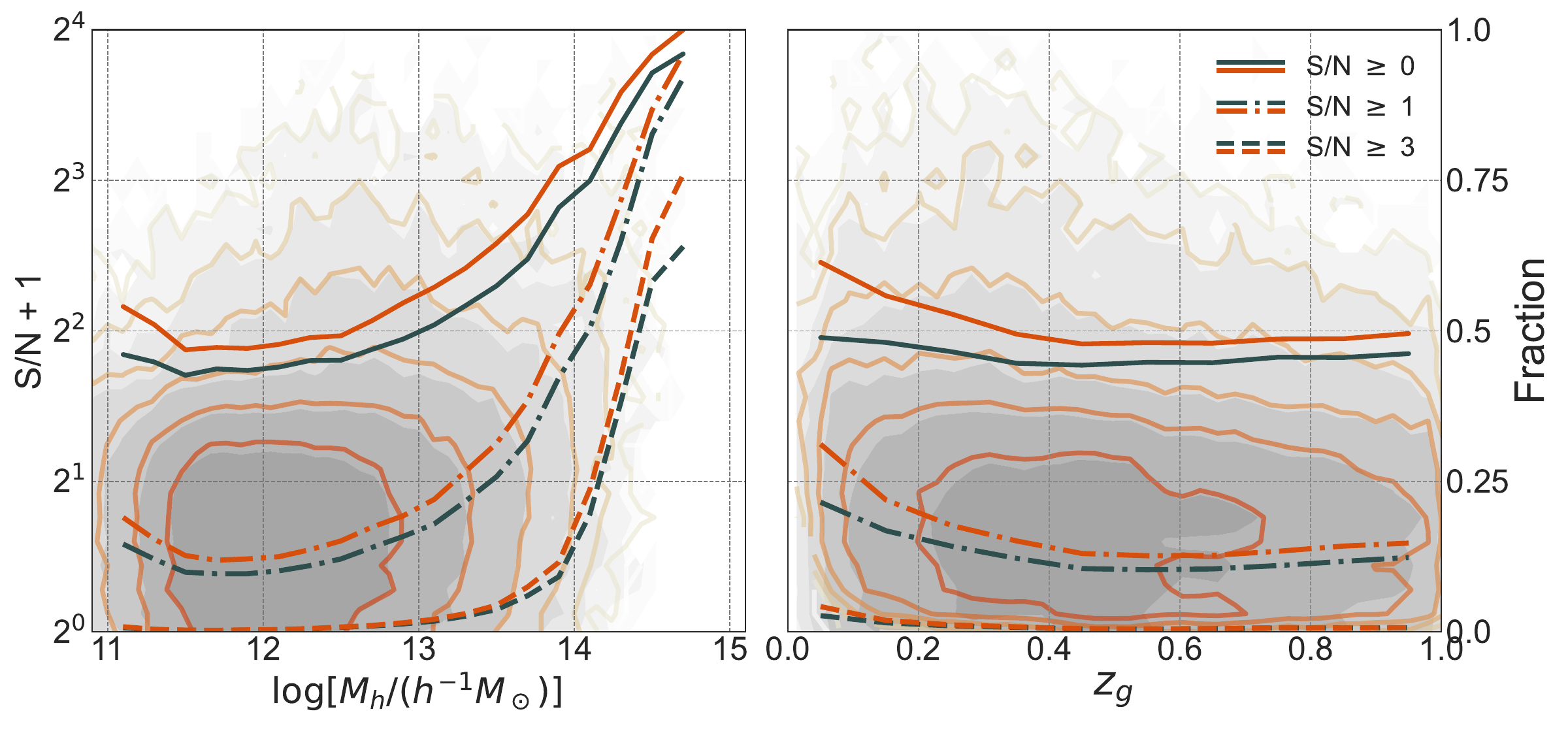}
  \caption{The contours outline the number density distribution for the $\rm{S}/\rm{N}$ ratios of X-ray groups as a function of halo mass (left panel) and redshift (right panel), respectively. The solid, dashdot, and dotted lines represent the fraction of the groups with $\rm{S}/\rm{N} \geq 0$, $\rm{S}/\rm{N} \geq 1$, and $\rm{S}/\rm{N} \geq 3$ as a function of halo mass and redshift, respectively. }
  \label{fig:sn}
\end{figure*}

\subsection{Source Count Rate} \label{sec:cts}
\subsubsection{Mean Background Subtraction Algorithm}
When calculating the count rate for each group, we locate the eFEDS field centered on the X-ray center and mask out all the contaminants that are not part of that group. Next, we set out to determine the X-ray background for each source. In S08 and W14, they determine the X-ray background for each group from an annulus with inner radius $R_{180}$ and a width of a few arcmins. Due to the relatively lower exposure time for eFEDS field and the fluctuation of the background estimate for each group is quite large, we determine the average count rate within the annuli at $1.5 R_{180} < R < 2.0 R_{180}$ for each galaxy group instead of the neighboring background subtraction algorithm. The average background count rate density is $\rho_{\rm bkg}^{\rm mean} \simeq 2.47 \times 10^{-5}$ cts/s/pixel$^2$. Besides the instrumental background, a fair proportion of the background photons might be emitted from the other extra-galactic sources, the galactic column density of neutral hydrogen ($n_{\rm H}$) might increase the fluctuations for the mean background level. Indeed, the $n_{\rm H}$ for the groups used in this work are varied from $\log\left(n_{\rm H}/\rm{cm}^{-2}\right) \simeq 20.26$ to $20.67$ given by HEALPIX resampling of Leiden/Argentine/Bonn Survey of Galactic HI \citep[][]{Kalberla..2005}. In Appendix~\ref{sec:model}, we show the energy conversion factor (ECF) that convert the soft X-ray band flux to the $0.2-2.3$ keV band count rates based on the power-law model but with $\log\left(n_{\rm H}/\rm{cm}^{-2}\right)$ ranged from $20.2$ to $20.7$, and their differences are relatively small ($\lesssim 0.1$ dex). Thus, we ignore the fluctuation in the estimate for $\rho_{\rm bkg}^{\rm mean}$. By subtracting the background counts scaled to the aperture radius of $R_{\rm X} = 0.5 R_{180}$, one can obtain the source count rates for each individual group. 

\subsubsection{Patrol Background Subtraction Algorithm}
In addition, we perform an alternative method to calculate the source count rates for each DESI group. First, we mask out all of the pixels that lie in at least one of the following regions:
\begin{enumerate}
  \item The regions that are not overlaid on the DESI footprint ($|b| \le 25^\circ$).
  \item The masked regions due to bright stars, globular clusters, or bad pixels in DESI footprint. 
  \item The regions enclosing the blind-detected sources.
  \item The pixels lie in the aperture radius of at least one DESI groups.
\end{enumerate}

If we set the aperture radius of each group to $R_{180}$, the patrol area makes less than $\sim$ 0.5 deg$^2$ of the total surveyed area. As discussed in the last section, the average count rates are concentrated within $R_{\rm X} = 0.5 R_{180}$. Therefore, we adopt the value from 0.5 to 1.0$R_{180}$ and derive the background count rate density $\rho_{\rm bkg}$ and the corresponding patrol area $A_{\rm bkg}$ as shown in figure~\ref{fig:patrol}. It can be seen from figure~\ref{fig:patrol} that the patrol background level is lower than the mean background level due to the contribution from these galaxy groups. The patrol background level show little dependence on the selection of aperture radius, the error is also very small when we adopt the value of $R/R_{180} = 1.0$. Therefore, we take use of the value of patrol background count rate density, $\rho_{\rm bkg}^{\rm ptrl} \simeq 2.31 \times 10^{-5}$ cts/s/pixel$^2$, when $R/R_{180} = 1.0$.

The patrol background count rates are mainly emitted from various sources such as instrumental background, local hot bubble, X-ray binaries, very distant ($z_g \ge 1$) groups and AGNs that are not resolved in eFEDS map. After removing the signals from these sources, the remaining are in principle emitted from the galaxy groups in the catalog used in this work only. However, the X-ray estimate might be impacted by the projection effect along the line-of-sight, causing the X-ray flux to be overestimated \citep{WangLei..2014}. For each group, one need to disentangle the count rates within $R_{\rm X}$ for each individual group.

Here we use a Monte Carlo mock to quantify the group X-ray luminosity overestimation due to the projection effect. Starting from all the groups in our sample, we first assume an average $L_{\rm X}-M_{h}$ relation to assign X-ray luminosities, $L_{\rm X, ass}$, to individual groups\footnote{As shown in section~\ref{sec:general}, we see that the redshift dependency is weak. We thus donot consider any redshift dependency in $L_{\rm X}-M_{h}$ relation.}. The initial guess is adopted from the fitting for the results using mean background subtraction algorithm. Then we convert the $L_{\rm X, ass}$ to the photon counts with an exposure time of 5 million seconds, which is by a factor of $\sim 4000$ longer than observation to ensure the signals for faint groups are sufficiently high. For each group, the mock photons are randomly generated following the $\beta-$model profile. After generating the mock image, we can derive the assigned ($N_{\rm ass}$) and projected ($N_{\rm pro}$) counts within a radius of $R_{\rm X}$ for each group in the same way as we did in observation. We calculate the ratio of the obtained $N_{\rm pro}$ and the assigned $N_{\rm ass}$, 
\begin{equation}
f_{\rm corr} = \cfrac{N_{\rm ass}}{N_{\rm pro}} ,
\end{equation}
which is used as the correction factor for each of our group. We use this factor to calculate the X-ray luminosity based on this algorithm for each group (see section~\ref{sec:xlums} for details) and obtain the tentative $L_{\rm X}-M_h$ relation (see section~\ref{sec:general} for details). Then we use this tentative relation to reproduce the above process, until there is no further change in the average $L_{\rm X}-M_{h}$ relation. In the final version, the typical value of $f_{\rm corr}$ is $\sim$ 0.31. In figure~\ref{fig:comparison}, we show the number density for $f_{\rm corr}$ as a function of $M_h$. As can be seen, small groups are heavily affected by the projection effect. However, the lower background level ($\rho_{\rm bkg}^{\rm ptrl} < \rho_{\rm bkg}^{\rm mean}$) also raises the flux estimate before corrected by projection effect. Both effects raise the scatter of the the $L_{\rm X}$ estimate.

\subsubsection{The $\rm{S}/\rm{N}$ ratios for individual group}

For each galaxy group, their signal-to-noise, $\rm{S}/\rm{N}$, is calculated using 
\begin{equation}
{\rm S}/{\rm N} = \frac{N_{\rm src}}{\sqrt{N_{\rm src} + N_{\rm bkg}}},
\end{equation}
where $N_{\rm bkg} = \pi R_{\rm X}^{2} \cdot \rho_{\rm bkg} t_{\rm exp}$ is the background photon counts scaled to the aperture of radius $R_{\rm X}$, $t_{\rm exp}$ is the mean exposure time of that source, and $N_{\rm src}$ is the net source photon counts within the aperture of radius $R_{\rm X}$\footnote{During the commission, light leak contamination was reported in TM5 and TM7 \citep{Predehl..2021}, which will affect the X-ray events with energies below $\sim 0.8$ keV. However, as we have tested by including or excluding the X-ray events detected by TM5 and TM7, the count rate of each group are in good agreement with each other. Thus in order to have higher $\rm{S}/\rm{N}$, TM5 and TM7 are kept in our analysis. }. Note that the $\rm{S}/\rm{N}$ derived based on patrol background subtraction algorithm is slightly higher than mean background subtraction algorithm.

In Figure~\ref{fig:sn}, we show the number density distribution for the $\rm{S}/\rm{N}$ ratios of X-ray groups based on different algorithm as a function of $M_h$ and $z_g$, respectively. For reference, we also plot the fraction of the groups above different $\rm{S}/\rm{N}$ thresholds as a function of $M_h$ and $z_g$, respectively. Among all the groups, there are $\sim 0.9 \%$ (5284, within which 4195 are in the blind detection source list) have $\rm{S}/\rm{N} \geq 3$ , $\sim 14.3 \%$ (84642) have $\rm{S}/\rm{N} \geq 1$, and $\sim 47.3 \%$ (278985) have $\rm{S}/\rm{N} \geq 0$ if we use the mean background subtraction algorithm, while $\sim 1.0 \%$ (6075, within which 4637 are in the blind detection source list) have $\rm{S}/\rm{N} \geq 3$ , $\sim 17.3 \%$ (102032) have $\rm{S}/\rm{N} \geq 1$, and $\sim 52.7 \%$ (311120) have $\rm{S}/\rm{N} \geq 0$ if we use the patrol background subtraction algorithm. The average $\rm{S}/\rm{N}$ is mainly lowered by small groups because the group X-ray luminosity positively correlate to their $M_h$. A little more than half of the small groups with $M_h \lesssim 10^{13}h^{-1}M_{\odot}$ have $N_{\rm src} < 0$ because of the negative expected median value for a source with nearly zero count rates relative to the background level.

\begin{figure}
    \centering
    \includegraphics[width=1\hsize]{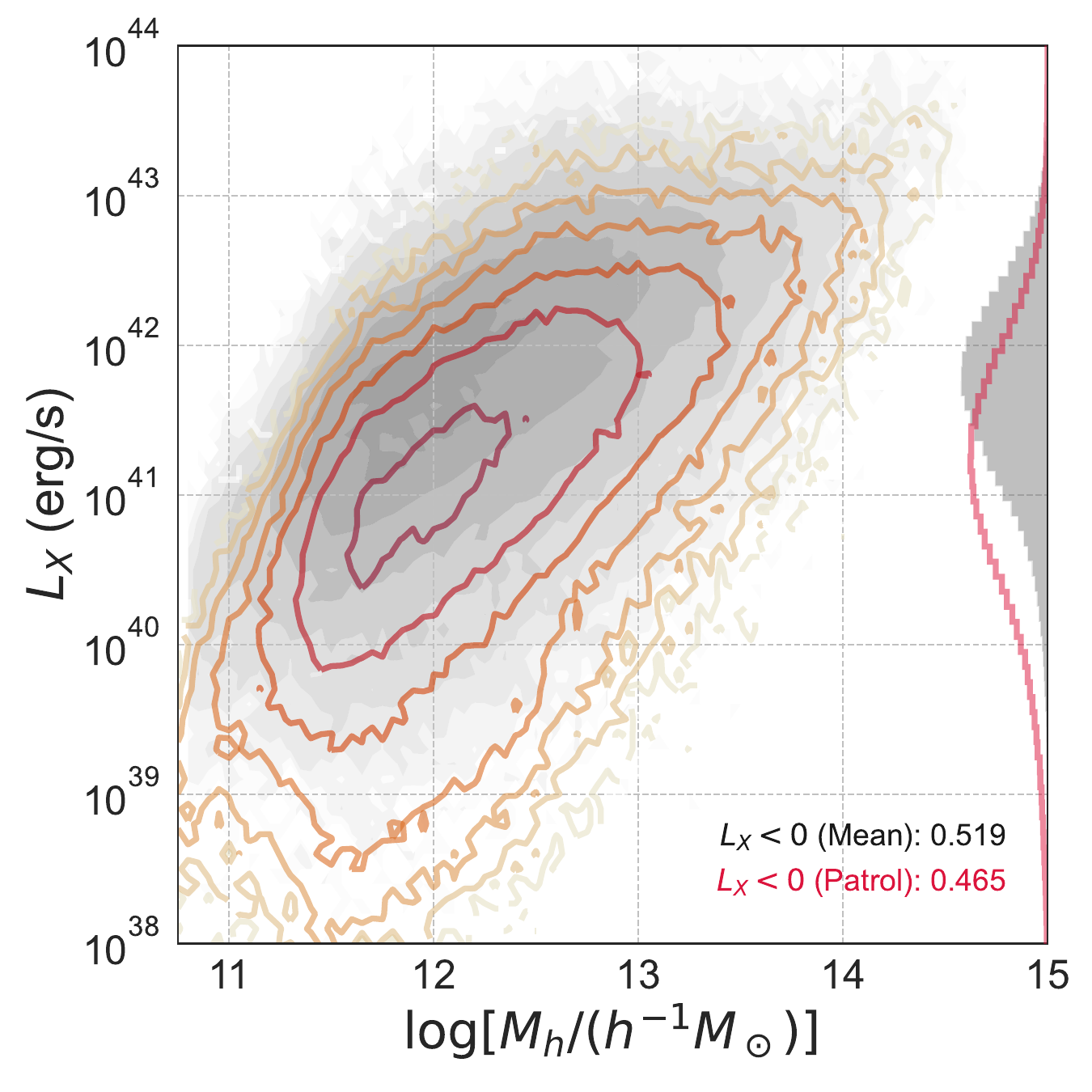}
    \caption{The grey filled contour shows the results given by mean background subtraction algorithm, while the red opened contour represents the results given by patrol background subtraction algorithm. The right side show the distributions of the $L_{\rm X}$ obtained by mean (grey shaded) and patrol (red opened) background subtraction algorithms, respectively. In the lower-right corner, we show the fration of the groups with $L_{\rm X} < 0$ obtained by both algorithms, respectively.}
    \label{fig:distributions}
\end{figure}

\begin{figure}
    \centering
    \includegraphics[width=1\hsize]{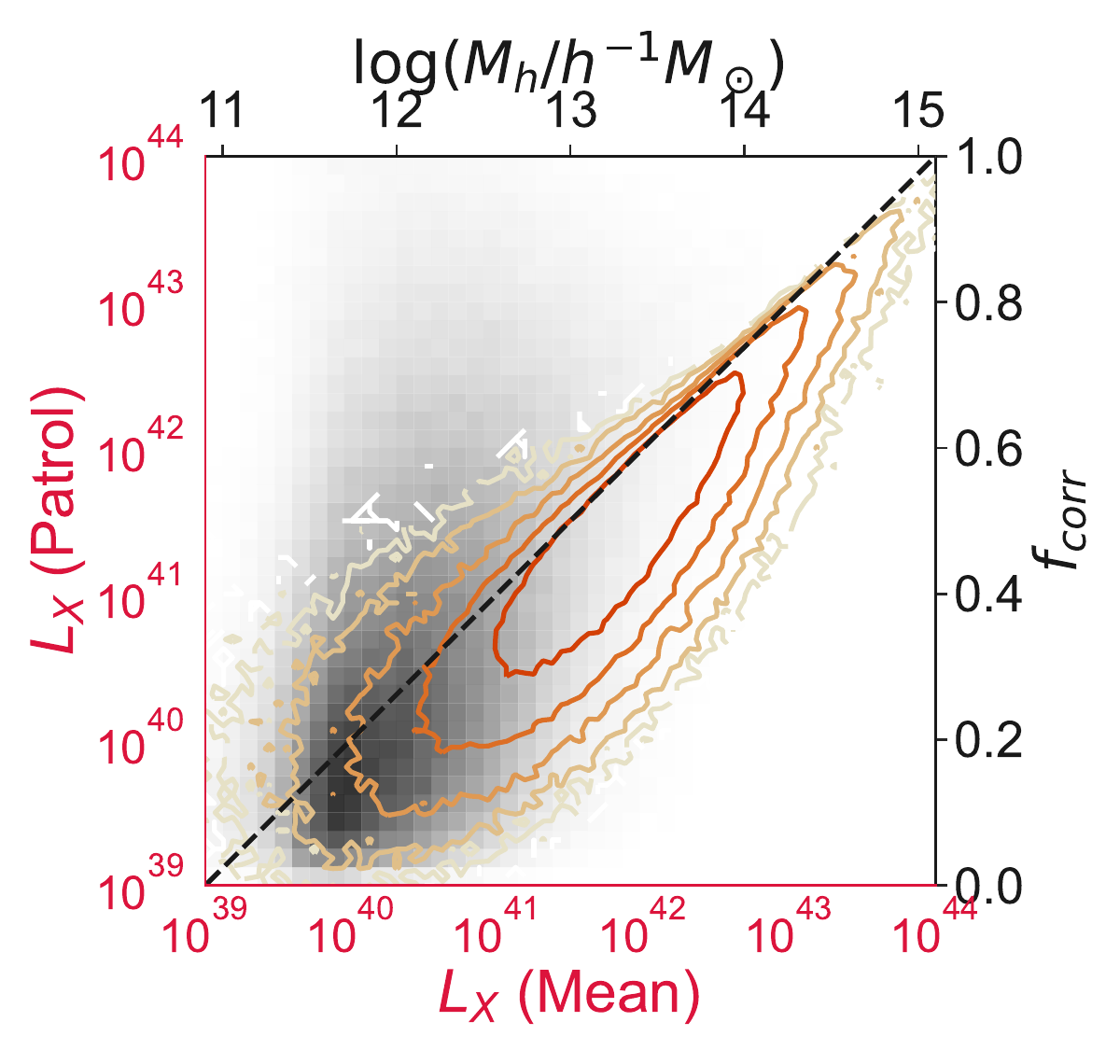}
    \caption{The grey map shows the number density distribution for $f_{\rm corr}$ as a function of $M_h$. The red contour represents the comparison between the $L_{\rm X}$ obtained using mean and patrol background algorithms. The dashed line is one-to-one correspondence between them. }
    \label{fig:comparison}
\end{figure}

\subsection{The X-ray Luminosities} \label{sec:xlums}

After deriving the count rates for each DESI group, we convert it into soft X-ray flux by dividing the source count rates, $C_{\rm src}$, to ECF. Assuming a spectral model, the ECF is obtained as the ratio of the count rate given by an \texttt{XSpec} mock spectrum to its model flux. The Ancillary Response File (ARF) and Response Matrix File (RMF), which are created for the mock spectrum, are generated by \texttt{eSASS} tool \texttt{srctool}\footnote{In practice, we un-correct the ARF by dividing the ``SPECRESP" by the correction ``CORRCOMB" when multiplying the model spectrum by the effective area \citep{LiuT..2022}.}. In this work, we adopt the ECF based on the power-law models with photon index $\Gamma = 2.0$. The details for our chosen are provided in Appendix~\ref{sec:model}. For an individual group, the X-ray luminosity can be expressed as
\begin{equation}
L_{\rm X} = \cfrac{4 \pi d_{L}^2 f_{\beta} \cdot C_{\rm src}}{g \left(n_{\rm H}, z_g, T\right)},
\end{equation}
and the source count rates can be expressed as
\begin{equation}
C_{\rm src} = \cfrac{f_{\rm corr} N_{\rm src}}{t_{\rm exp}},
\end{equation}
where $d_{L}$ is the luminosity distance of the group, $f_{\beta}$ is the extension correction factor, $f_{\rm corr}$ is the flux fraction of an X-ray group in a multi-cluster detection, $N_{\rm src}$ is the source count rates within aperture of radius $R_{\rm X}$, $t_{\rm exp}$ is the exposure time, and $g \left(n_{\rm H}, z_g, T\right)$ is the ECF depend on column density of neural hydrogen ($n_{\rm H}$), redshift ($z_g$), and temperature ($T$). In this work, we adopt the $n_{\rm H}$ given by HEALPIX resampling of Leiden/Argentine/Bonn Survey of Galactic HI \citep[][]{Kalberla..2005}. We note that the correction factor are set to $f_{\rm corr} = 1$ for the results obtained using mean background subtraction algorithm.

In figure~\ref{fig:distributions}, we compare the two sets of $L_{\rm X}$, those obtained by mean background subtraction algorithm are generally higher than that obtained by patrol background subtraction algorithm at positive $L_{\rm X}$, but the latter has more positive values than the former. Although $\sim 50 \%$ groups have negative source count rate and their $L_{\rm X}$ are negative as a result, we retain all of the samples in subsequent analysis. In figure~\ref{fig:comparison}, we show the comparison between the $L_{\rm X}$ obtained by mean and patrol background algorithm. The difference tend to be larger at the lower $L_{\rm X}$ end due to the projection correction. 

\begin{figure}
\centering
    \includegraphics[width=1\hsize]{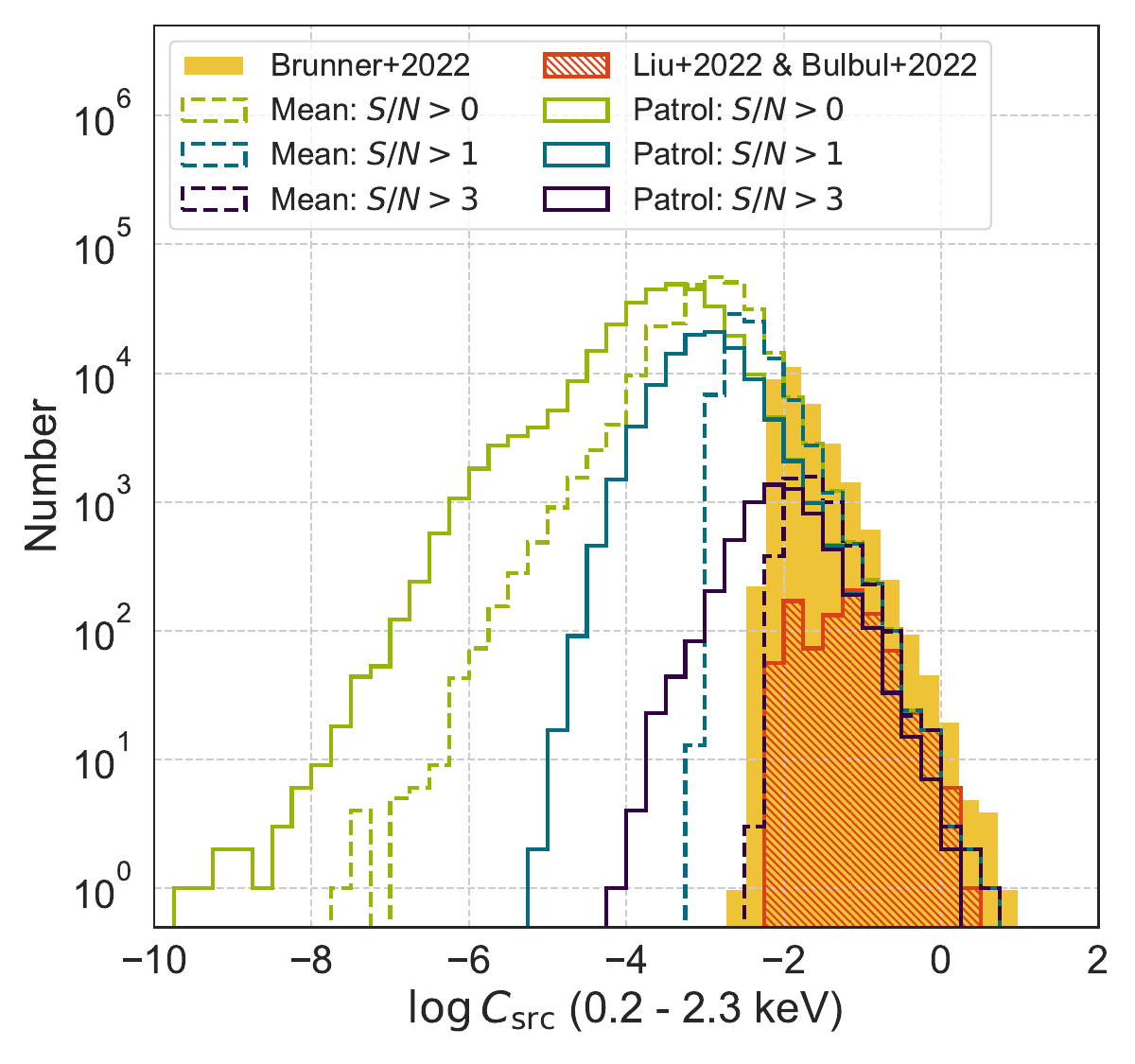}
  \caption{The $0.2-2.3$ keV band source count rate, $C_{\rm src}$, distribution for the results using mean (dashed) and patrol (solid) background subtraction algorithms for $\rm{S}/\rm{N} > 0$ (green), $\rm{S}/\rm{N} \geq 1$ (blue), and $\rm{S}/\rm{N} \geq 3$ (purple) groups, respectively. The yellow filled histogram shows the results for all the blind-detected sources based on eFEDS \citep[][]{Brunner..2022}, while the red hatched histogram shows the group candidates filtered by \citet{LiuA..2022} and \citet{Bulbul..2022} and overlaid on DESI footprint.}
  \label{fig:cts}
\end{figure}

\begin{figure*}
\centering
    \includegraphics[width=0.95\hsize]{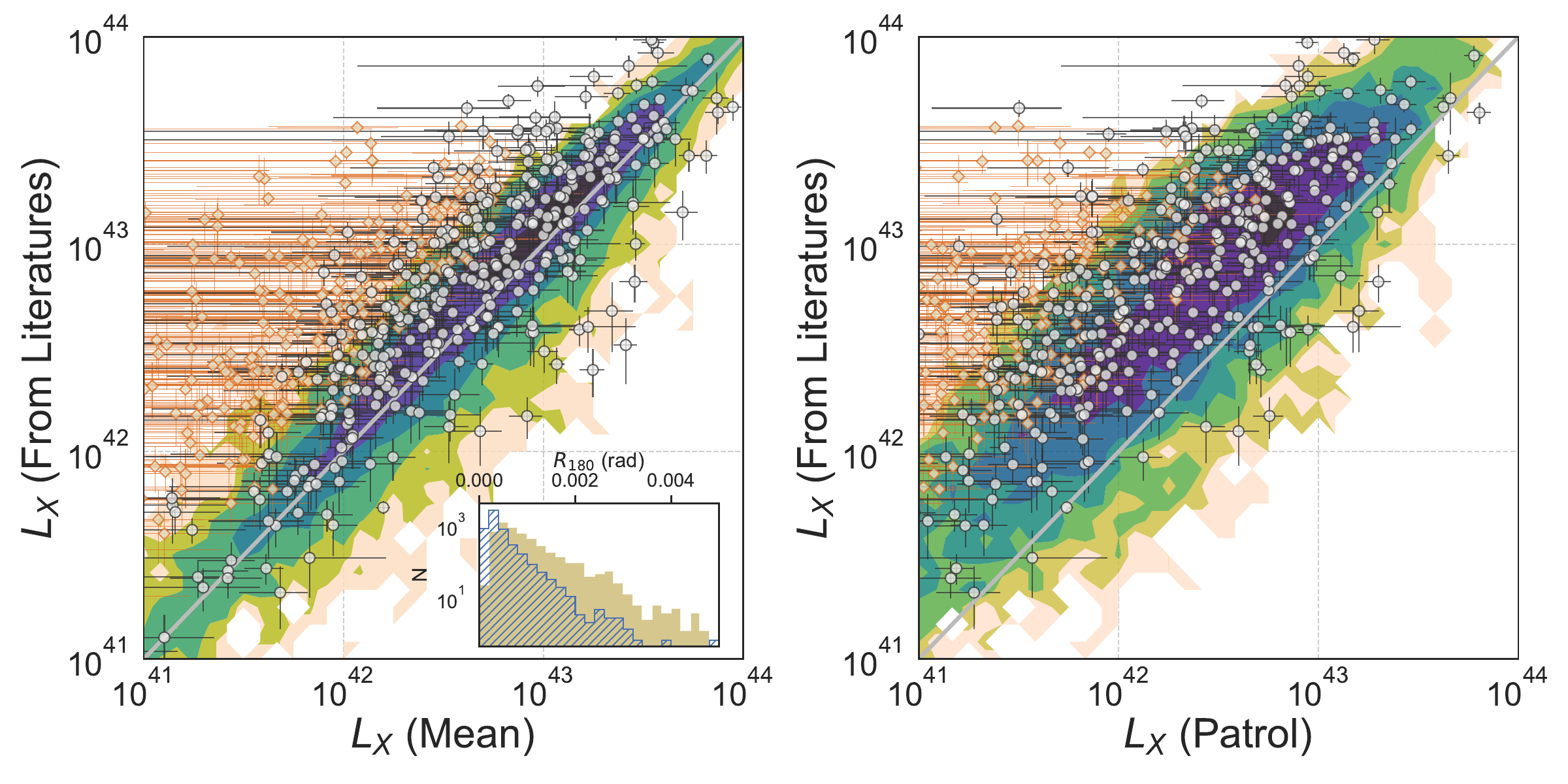}
  \caption{The comparison of the $L_{\rm X}$ obtained using mean (left) and patrol (right) background subtraction algorithms with the $L_{\rm X}$ taken from the literatures: 1). \citet[][]{Brunner..2022}: contour map; 2). \citet[][]{Ranalli..2015}: symbols with error bars, the groups brighter and fainter than $\sim 2 \times 10^{-15}$ erg/s/cm$^2$ are shown in circle and diamond, respectively. The grey solid lines in both panels are one-to-one correspondence between the results to be compared. The plot in the inset in the lower-right of the left panel shows the $R_{180}$ distributions for the groups with $L_{\rm X}$ derived using mean background subtraction algorithm lower (hatched) and higher (filled) than the $L_{\rm X}$ given by \citet[][]{Brunner..2022}. Note that the $L_{\rm X}$ taken from the literatures have been converted using the ECFs given by this study (see section~\ref{sec:comparison}).}
  \label{fig:previous}
\end{figure*}

\subsection{Comparison with existing X-ray Clusters} \label{sec:comparison}

Having obtained the X-ray luminosity for all of our groups, we proceed to compare our X-ray measurements with the results that available in previous studies. The datasets we perform the cross-identification are as follows:
\begin{enumerate}
\item eFEDS X-Ray Catalog \citep{Brunner..2022}: A catalog of blind-detected sources based on eFEDS. This catalog contains $\sim 33000$ blind-detected sources, including 542 extended sources that are regarded as groups \citep{LiuA..2022} and 346 point sources but suspected as groups in disguise \citep{Bulbul..2022}. The count rate for each source has been PSF-corrected. We compare the results for the 10932 DESI groups hosting resolved X-ray with the counterparts in this catalog only. To show a fair comparison, the corresponding rest-frame $0.1-2.4$ keV band X-ray luminosity are converted using the ECFs given by this work.
\item XMM-ATLAS Survey \citep{Ranalli..2015}: XMM-Newton observations in the H-ATLAS SDP area, covering $\sim 7$ deg$^{2}$ with a limit of $\sim 2 \times 10^{-15}$ erg/s/cm$^2$ in $0.5-2.0$ keV band and overlapping with the eFEDS footprints. This catalog gives the observed $0.5-2.0$ keV band flux of each source by assuming a power-law spectra with a photon index of $\Gamma = 1.7$ and Galactic absorption of $n_{\rm H} = 2.3 \times 10^{20}$ cm$^{-2}$. We cross-match the DESI group catalog with XMM-ATLAS samples within a tolerance of 20 arcsec. There are 961 DESI groups have counterparts in their catalog, 409 of them have flux larger than $\sim 2 \times 10^{-15}$ erg/s/cm$^2$ in eFEDS observations. Note that this catalog does not give the redshift of each XMM-ATLAS source. In order to make a fair comparison, the redshifts of those match XMM-ATLAS sources are assigned from their counterparts in our sample, and we convert the flux to the rest-frame $0.1-2.4$ keV band flux corrected for Galactic absorption. 
\end{enumerate}

Figure~\ref{fig:cts} displays the $0.2-2.3$ keV band $C_{\rm src}$ distributions for DESI groups and blind-detected sources. Compared to the X-ray groups detected by \citet{LiuA..2022} and \citet{Bulbul..2022} which are shown using the red hatched histogram, our X-ray groups with S/N$\ge 3$ (purple solid and dashed histograms) are about an order of magnitude more.  The shift of the $C_{\rm src}$ given by the patrol background subtraction algorithm in the $x$-axis direction is mainly caused by the projection correction factor, which is evident by comparing with the dashed histogram for the results using the mean subtraction algorithm. Although the vast majority of our groups do not have S/N$\ge 3$ X-ray detection, they can still be used to carry out scientific studies, e.g., through stacking algorithm, etc.  

In figure~\ref{fig:previous}, we show the comparison of the X-ray luminosities between our measurements and those obtained from literatures. First, our results obtained by mean background subtraction algorithm are slightly lower ($\lesssim 0.05$ dex) than that given by \citet{Brunner..2022}, which might due to the selection of the aperture radius. The inset in the lower-right of the left panel show the $R_{180}$ distributions for the groups with $L_{\rm X}$ derived using mean background subtraction algorithm lower (hatched) and higher (filled) than that given by \citet{Brunner..2022}, respectively. Clearly, the former is generally smaller than the latter, implying the selection of the aperture radius might affect the results. Because of the count rate is integrated to the aperture radius, the X-ray luminosities of groups with large radius in projection are generally overestimated and vice versa. In addition, our results are systematically lower than those obtained by \citet{Ranalli..2015} and the $\rm{S}/\rm{N}$ of our results are generally lower because the average exposure time of eFEDS is lower than that of XMM-Newton. However, X-ray selected samples are known to miss galaxy groups with lower X-ray flux. We separate the groups matched by \citet{Ranalli..2015} into those with X-ray flux brighter and fainter than $\sim 2 \times 10^{-15}$ erg/s/cm$^2$ in eFEDS observations, those above the flux threshold show good agreement with \citet{Ranalli..2015}. From the right panel of figure~\ref{fig:previous}, we see that our results obtained by patrol background subtraction algorithm are  systematically lower ($\sim 0.15$ dex) than that of \citet{Brunner..2022} because we corrected for the projection effect. Taking into account with and without the projection effect, our X-ray measurements for the corresponding groups are in nice agreement with both of these studies.

\section{X-ray luminosity - halo mass relation}\label{sec:general}

One of the most important X-ray scaling relations for cosmology with galaxy groups is the $L_{\rm{X}}-M_h$ relation. To derive the $L_{\rm{X}}-M_h$ relation, a complete sample are required because the scatter of $L_{\rm{X}}$ is quite large at a given $M_h$ and a X-ray flux-limited sample suffers from the selection bias that brighter objects can be observed out to farther distance. Such bias has previously been taken into account for obtaining $L_{\rm{X}}-M_h$ relations based on X-ray selected group sample using different assumptions \citep[e.g.,][]{Vikhlinin..2009, Pratt..2009, Mittal..2011, Lovisari..2015}.

Now that we have measured the X-ray luminosities for {\it all} the DESI groups overlaid on the eFEDS footprints, i.e., the X-ray measurements are completed for the groups at given $M_h$ and $z_g$. It is thus quite straight forward to derive the related $L_{\rm{X}}-M_h$ relations. We use the following two ways to derive the relations and make self-consistent checks.

\subsection{Stacking Method}

In order to check if there are any redshift dependence in the $L_{\rm{X}}-M_h$ relations, we first separate the groups into different $M_h$ and $z_g$ bins. To get sufficient signals for our investigation, we stack the X-ray luminosities for groups in each bin. The stacked X-ray luminosity $L_{\rm X,S}$ for given $N$ groups can be obtained in two ways. The first way is calculating the mean $L_{\rm X}$ directly: $L_{\rm X,S} = \sum^{N}_{i=1} \frac{L_{{\rm X}, i}}{N}$, and the second way to calculate the stacked X-ray luminosity can be expressed as 
\begin{equation}\label{eq:5}
L_{\rm X,S} = f_{\beta} \cdot \frac{\sum\limits^{N}_{i=1} N_{{\rm src}, i} }{\sum\limits^{N}_{i=1}\frac{{g}_i \cdot t_{{\rm exp},i}}{4\pi d_{L,i}^2 \cdot f_{{\rm corr}, i}}} ,
\end{equation}
where $d_{L,i}$ is the luminosity distance for i$th$ group. Both calculations are nearly consistent, and we take use of the results given by Equation~\ref{eq:5} unless stated otherwise.

\begin{figure}
\centering
    \includegraphics[width=1.\hsize]{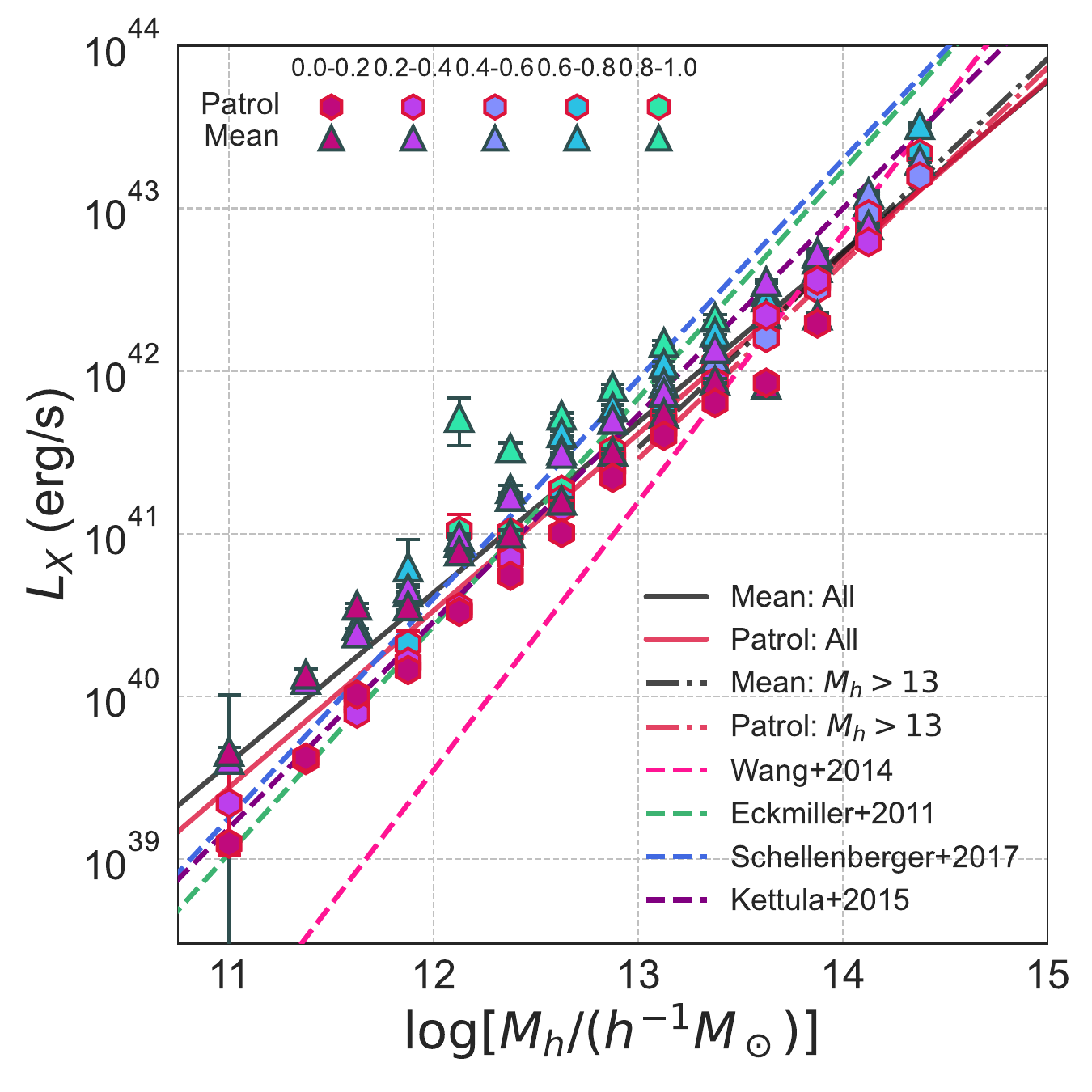}
  \caption{The X-ray group luminosity, $L_{\rm{X}}$, obtained by both algorithms versus halo mass, $M_h$, for all the DESI groups used in this work. The triangles and hexagons with error bars represent the {\it stacked} X-ray luminosity, $L_{\rm X,S}$, as a function of $M_h$ color-coded by redshift for the results obtained by both algorithms, respectively. Only the data bins with at least 50 groups are plotted. The solid lines shows our best-fit for the results of overall samples, while the dash-dot lines are the best-fit for the $M_h \ge 10^{13}h^{-1}M_\odot$ subsamples. The magenta, green, blue, and purple dashed lines show the results obtained by \citet{WangLei..2014}, \citet{Eckmiller..2011}, \citet{Schellenberger.Reiprich2017}, and \citet{Kettula..2015}, respectively. }
  \label{fig:LX}
\end{figure}

In figure~\ref{fig:LX}, we show the stacked X-ray luminosity $L_{\rm X,S}$ obtained in different methods color-coded by their $z_g$. For the results obtained by the same method, their normalizations as well as the slopes show no significant differences. However, the stacked $L_{\rm X,S}$ for patrol background subtraction algorithm are slightly lower than that for mean background subtraction algorithm at low $M_h$ end. The projection effect tend to be more evident for small groups, the lower background level cannot fully compensate for the reduced flux due to flux correction factor.

\subsection{Direct model fitting}

The other way to obtain the $L_{\rm{X}}-M_h$ relation is by assuming a functional form and fit for the related parameters. Here we assume the $L_{\rm{X}}-M_h$ relation has a power law form:
\begin{equation}
\left(\frac{L_{\rm{X}}}{\rm{erg/s}}\right) = 10^{A}\cdot{\left(\frac{M_h}{h^{-1}M_\odot}\right)}^{B},
\end{equation}
where $10^{A}$ is the normalization and $B$ is the slope. Some previous studies \citep[e.g.,][]{Vikhlinin..2009b, Reichert..2011} have taken into account the redshift evolution of the normalization by multiplying $\left[\rm{H}(z)/\rm{H}_0\right]^{C}$, where $\rm{H}(z)$ is the Hubble-Lema\^{i}tre parameter, $\rm{H}_0$ is the Hubble constant and $C$ is a constant. However, as we have not seen any significant redshift evolution behavior in this study, we do not consider the redshift evolution term here.

Owing to the fact that the photon counts are very small for numerous groups, especially at the low-mass end, we model the $L_{\rm{X}}-M_h$ relation such that the observed $L_{\rm X}$ is distributed around the scaling relation in a Poisson form. The probability for the i$th$ group is given as
\begin{equation}
\mathscr{P}\left(L_{{\rm X},i}|M_{h,i},A,B\right) = \frac{e^{\lambda_i} \cdot \lambda_{i}^{\frac{{g}_i \left(\frac{L_{{\rm X},i}}{f_\beta} + L_{{\rm B},i} \right) \cdot t_i}{4\pi d_{\rm{L},i}^{2}}}}{\Gamma\left[\frac{{g}_i \left(\frac{L_{{\rm X},i}}{f_\beta} + L_{{\rm B},i} \right) \cdot t_i}{4\pi d_{\rm{L},i}^{2}} + 1\right]},
\end{equation}
where $L_{{\rm X},i}$, $M_{h,i}$, $f_{\beta}$, $t_i$, $d_{\rm{L},i}$, and $g_i$ are the X-ray luminosity, halo mass, $\beta$-profile extension correction, mean exposure time, luminosity distance, and ECF of the i$th$ group, respectively. Also, $L_{{\rm B},i}$ is the subtracted background luminosity scaled to the $R_{\rm X}$, which can be expressed as: $L_{{\rm B},i} = \rho_{\rm bkg} \cdot \pi R_{\rm X}^2 \cdot \cfrac{4\pi d_{\rm{L},i}^2}{g_i}$. Note that the term, $\cfrac{{g}_i \left(\frac{L_{{\rm X},i}}{f_\beta} + L_{{\rm B},i} \right) \cdot t_i}{4\pi d_{\rm{L},i}^{2}}$, in the denominator Gamma function, is the overall photon events within a radius of $R_{\rm X}$ for the i$th$ group. We assume that the X-ray luminosity for each group is determined by their $M_h$ only, each group has an expected photon events, $\lambda_i$, and it is defined as:
\begin{equation}
\lambda_i = \cfrac{{g}_i \left(\frac{\left<L_{{\rm X},i}\right>}{f_\beta} + L_{{\rm B},i} \right) \cdot t_i}{4\pi d_{\rm{L},i}^{2}},
\end{equation}
where $\left<L_{{\rm X},i}\right> = 10^{A}\cdot{\left(\frac{M_{h,i}}{h^{-1}M_\odot}\right)}^{B}$. This yields a likelihood function that can be written as $\ln \mathscr{L} \equiv \sum\limits^{N}_{i = 1} \ln \mathscr{P}\left(L_{{\rm X},i}|M_{h,i},A,B\right)$ and we need to find the best-fit parameters  that maximizes the likelihood.

\subsection{Results}
Our best-fit $L_{\rm{X}}-M_h$ relations for both algorithms are presented in figure~\ref{fig:LX}, where we report a normalize of $10^{28.46\pm0.03}$ with a slope of $1.024\pm0.002$ for mean background subtraction algorithm, and a normalize of $10^{26.73\pm0.04}$ with a slope of $1.140\pm0.003$ for patrol background subtraction algorithm. Very encouragingly, both results are consistent and show nice agreement with their $L_{\rm X,S}$, respectively, demonstrating that our model constraints are self-consistent. 

For comparison, we also plot the results obtained previously by \citet{WangLei..2014}, \citet{Eckmiller..2011}, \citet{Schellenberger.Reiprich2017}, and \citet{Kettula..2015} in Figure~\ref{fig:LX}. Note that \citet{Eckmiller..2011} and \citet{Kettula..2015} give the $L_{\rm{X}}-M_{180}$ relations and \citet{Schellenberger.Reiprich2017} gives the $L_{\rm{X}}-M_{500}$ scaling relation after correcting the Malmquist and Eddington biases. Their group mass indicators are slightly different from ours. To unify the definition of $M_h$, we convert the $M_{180}$ and $M_{500}$ to $M_h$ ($M_{180}$) by assuming that  the dark matter halos follow a Navarro-Frenk-White \citep[NFW,][]{NFW} density profile with concentration parameters given by the concentration-mass relation of \citet{Maccio..2007}. Based on this assumption, we get $M_{h}/M_{180} = 1.03$ and $M_{h}/M_{500} = 1.38$ when the concentration index is $c_{180} = 6$. Note that the concentration index is negatively correlate to the $M_h$, and the $M_{h}/M_{180}$ and $M_{h}/M_{500}$ are varied with concentration index. However, the difference between the $M_{h}/M_{180}$ ($M_{h}/M_{500}$) by adopting $c_{180} = 5$ and $c_{180} = 12$ are smaller than $\lesssim 0.01$ dex ($\lesssim 0.07$ dex), we thus ignore the change of the slope for these relations taken from the literature. 

Clearly, our model constrain of the slopes, $1.024 \sim 1.140$, are flatter than the slope ranging of $1.27 \sim 1.65$ obtained by the literature but close to the slope predicted by self-similar relation: $L_{\rm X}^{0.1-2.4{\rm keV}} \propto M$ (Equation 26 in \citealt{Schellenberger.Reiprich2017}). However, these previous results are generally obtained from the samples with $M_h \gtrsim 10^{13}h^{-1}M_{\odot}$, the slope of the $L_{\rm X}-M_h$ relation might be different in different $M_h$ ranges. Here we perform the same method to fit the $L_{\rm{X}}-M_{h}$ relation for groups with $M_h \ge 10^{13}h^{-1}M_{\odot}$, and we plot the best-fit results for both algorithms in dash-dot lines in figure~\ref{fig:LX}, where we report a normalize of $10^{26.91\pm0.06}$ with a slope of $1.135\pm0.004$ for mean background subtraction algorithm, and a normalize of $10^{25.64\pm0.08}$ with a slope of $1.217\pm0.005$ for patrol background subtraction algorithm. These results are still flatter than those taken from the literatures but steeper than the result obtained for overall samples. 

As pointed by \citet{Lovisari..2021}, a mass-dependent bias in the group mass estimate might potentially affect the slope of the $L_{\rm X}-M_h$ relation, especially at the low-mass end. Due to the fact that it is difficult to distinguish the low-temperature emitting gas of small groups from the galactic foreground, the X-ray properties of them are generally observed out to a smaller radial extent. An estimate of the group mass based on X-ray information and hydrostatic equilibrium might affect the shape of the $L_{\rm X}-M_h$ relation. In this work, the group mass, $M_h$, is obtained from the abundance matching according to the accumulative halo mass and group luminosity functions, and the uncertainty of $M_h$ is less than $\sim 0.4$ dex. Thus independently estimated $M_h$ might make the $L_{\rm X}-M_h$ relation less prone to be biased.

\section{Conclusion} \label{sec:conclusion}

In this study, using the optical information, such as the position of the massive galaxy members, $M_h$, and $z_g$, we used two different algorithms to measure the luminosities in soft X-ray (rest-frame $0.1 -2.4$ keV) band for $\sim 600,000$ groups identified from DESI DR9 and overlaid on the footprints of the eFEDS, ranging in redshifts of $0.0 \le z_g \le 1.0$ and group mass of $10^{10.76}h^{-1}M_{\odot} \le M_h \le 10^{15.0}h^{-1}M_{\odot}$.
The main results of this paper are summarized as follows.

\begin{enumerate}

\item Among these groups, $\sim 0.9 \%$ of them have ${\rm S}/{\rm N} \ge 3$, $\sim 14.3 \%$ of them have ${\rm S}/{\rm N} \ge 1$, and $\sim 47.3 \%$ of them have ${\rm S}/{\rm N} > 0$ when we subtract the background using the average count rate density in the background ring of each group, while the percentages are slightly higher ($\sim 1.0 \%$, $\sim 17.3 \%$, and $\sim 52.7 \%$ for ${\rm S}/{\rm N} \ge 3$, ${\rm S}/{\rm N} \ge 1$, and ${\rm S}/{\rm N} > 0$, respectively) when we subtract the background using the count rate density averaged over the regions that not lie within $R_{180}$ of any groups. By comparing to the blind-detected X-ray groups based on eFEDS, the number of X-ray groups been detected with  ${\rm S}/{\rm N} \ge 3$  have increased nearly by a factor of 6.
 
\item  By stacking the X-ray images of the groups that have no resolved X-ray centers in different $M_h$ and $z_g$ bins. The BGG can well represent the X-ray peak of a group system, and the average surface brightness profiles roughly follow the $\beta-$model prediction. We measure the  stacked X-ray lumonosities around similar mass groups that are divided into five redshift bins. We find the X-ray luminosity scales linearly with halo mass and is independent of the redshift.

\item By properly taking into account the Poisson fluctuations, we obtain the overall scaling relations between X-ray luminosity and halo mass mass with $L_{\rm X} = 10^{28.46 \pm 0.03}M_h^{1.024\pm0.002}$ and $L_{\rm X} = 10^{26.73 \pm 0.04}M_h^{1.140\pm0.003}$ based on the results using two different algorithms, both of which are consistent with the results obtained using stacking method. Both scaling relations are flatter than those obtained previously by \citet{WangLei..2014}, \citet{Eckmiller..2011}, \citet{Schellenberger.Reiprich2017}, and \citet{Kettula..2015}, but closer to the self-similar prediction.

\end{enumerate}

Combined with the DESI Legacy Imaging Surveys, our results display the capability of eROSITA to determine the X-ray emission out to $R_{180}$ for a deep flux limited galaxy group sample. Future analysis using eROSITA all-sky survey data, combined with the group catalog with more accurate redshifts, would provide much enhanced quantitative X-ray measurement. Detailed analysis of the hot gas evolution in galaxy groups, and the physical modeling of their evolution will be presented in  forthcoming papers.

\section*{Acknowledgements}
We are thankful for Teng Liu for helpful discussions. This work is supported by the National Science Foundation of China (Nos. 11833005, 11890692, 11621303, 12141302), 111 project No. B20019, and Shanghai Natural Science Foundation, grant No.19ZR1466800. We acknowledge the science research grants from the China Manned Space Project with No.CMS-CSST-2021-A02.
The computations in this paper were run on the Gravity Supercomputer at Shanghai Jiao Tong University.

This work is based on data from the DESI Legacy Imaging Surveys. The DESI Legacy Imaging Surveys consist of three individual and complementary projects: the Dark Energy Camera Legacy Survey (DECaLS), the Beijing-Arizona Sky Survey (BASS), and the Mayall z-band Legacy Survey (MzLS). DECaLS, BASS and MzLS together include data obtained, respectively, at the Blanco telescope, Cerro Tololo Inter-American Observatory, NSF’s NOIRLab; the Bok telescope, Steward Observatory, University of Arizona; and the Mayall telescope, Kitt Peak National Observatory, NOIRLab. NOIRLab is operated by the Association of Universities for Research in Astronomy (AURA) under a cooperative agreement with the National Science Foundation. Pipeline processing and analyses of the data were supported by NOIRLab and the Lawrence Berkeley National Laboratory (LBNL). Legacy Surveys also uses data products from the Near-Earth Object Wide-field Infrared Survey Explorer (NEOWISE), a project of the Jet Propulsion Laboratory/California Institute of Technology, funded by the National Aeronautics and Space Administration. Legacy Surveys was supported by: the Director, Office of Science, Office of High Energy Physics of the U.S. Department of Energy; the National Energy Research Scientific Computing Center, a DOE Office of Science User Facility; the U.S. National Science Foundation, Division of Astronomical Sciences; the National Astronomical Observatories of China, the Chinese Academy of Sciences and the Chinese National Natural Science Foundation. LBNL is managed by the Regents of the University of California under contract to the U.S. Department of Energy. The Photometric Redshifts for the Legacy Surveys (PRLS) catalog used in this paper was produced thanks to funding from the U.S. Department of Energy Office of Science, Office of High Energy Physics via grant DE-SC0007914.

This work is also based on data from eROSITA, the soft X-ray instrument aboard SRG, a joint Russian-German science mission supported by the Russian Space Agency (Roskosmos), in the interests of the Russian Academy of Sciences represented by its Space Research Institute (IKI), and the Deutsches Zentrum f{\"u}r Luftund Raumfahrt (DLR). The SRG spacecraft was built by Lavochkin Association (NPOL) and its subcontractors, and is operated by NPOL with support from the Max Planck Institute for Extraterrestrial Physics (MPE). The development and construction of the eROSITA X-ray instrument was led by MPE, with contributions from the Dr. Karl Remeis Observatory Bamberg \& ECAP (FAU Erlangen-Nuernberg), the University of Hamburg Observatory, the Leibniz Institute for Astrophysics Potsdam (AIP), and the Institute for Astronomy and Astrophysics of the University of T{\"u}bingen, with the support of DLR and the Max Planck Society. The Argelander Institute for Astronomy of the University of Bonn and the Ludwig Maximilians Universit{\"a}t Munich also participated in the science preparation for eROSITA. The eROSITA data shown here were processed using the eSASS software system developed by the German eROSITA consortium.

\section*{Data Availability}
The data underlying this article will be shared on reasonable request to the corresponding author. The data underlying this article are avaliable at \url{https://github.com/Al-YL/XLumsForDESIGroups/tree/main/DR9Y1_zbd}.







\appendix

\section{Theoretical Models for X-ray Emission}\label{sec:model}

\begin{figure}
\centering
    \includegraphics[width=1.\hsize]{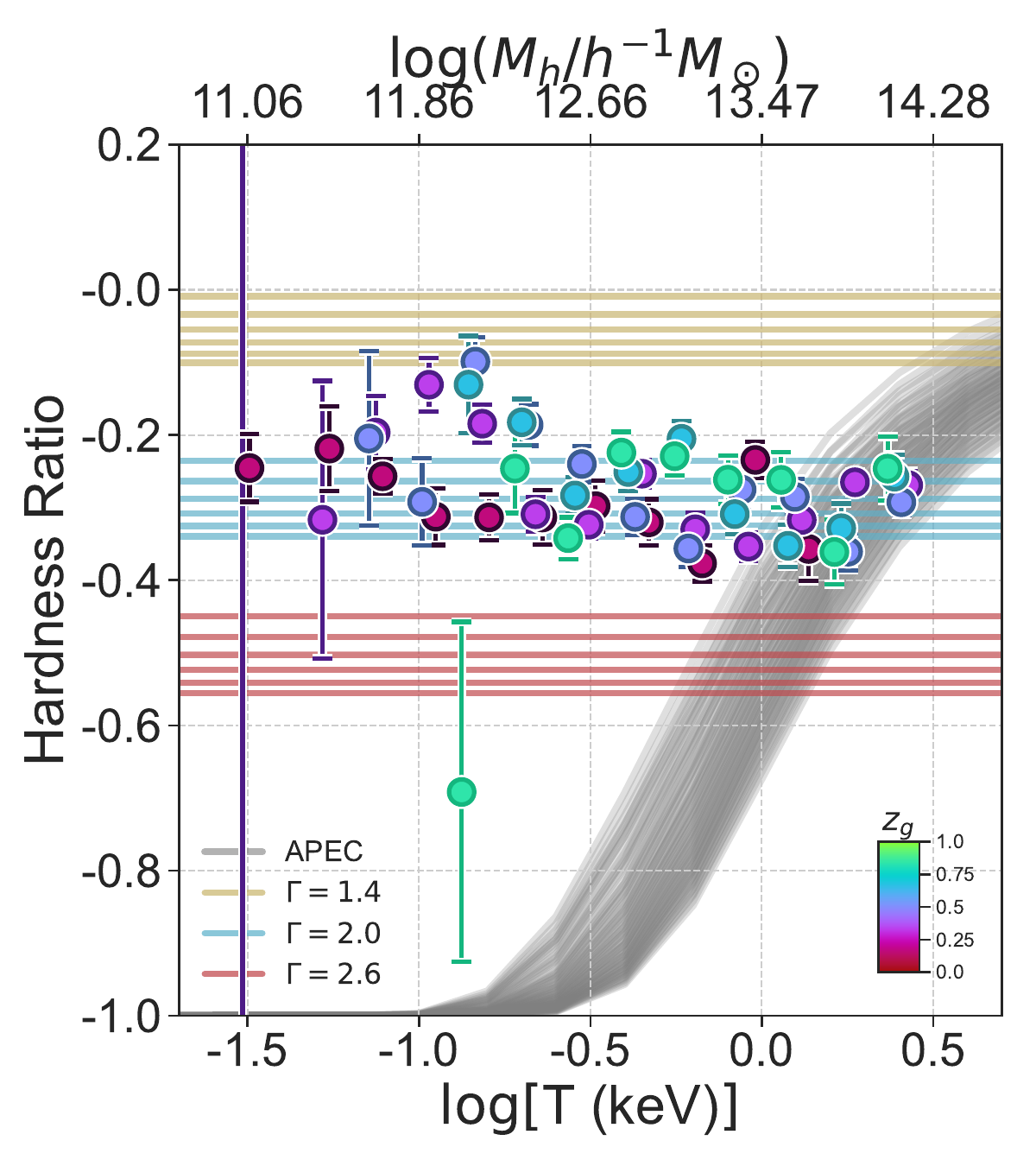}
  \caption{The hardness ratio as a function of gas temperature for groups at different redshifts. We manually offset the error bars on x-axis for clarity. Only the data bins with at least 100 groups are plotted. The grey lines are theoretical hardness ratio given by APEC models with different parameters (see details in appendix~\ref{sec:model})  taking into account the eFEDS effective area. The yellow, cyan, and red lines represent the theoretical hardness ratio given by power law models with $\Gamma=1.4$, $\Gamma=2.0$, and $\Gamma=2.6$, respectively. The $M_h$ corresponding to gas temperature are derived from the $M_{h}-T$ scaling relation \citep{Babyk..2023}.}
  \label{fig:hardness}
\end{figure}

\begin{figure}
\centering
    \includegraphics[width=1.\hsize]{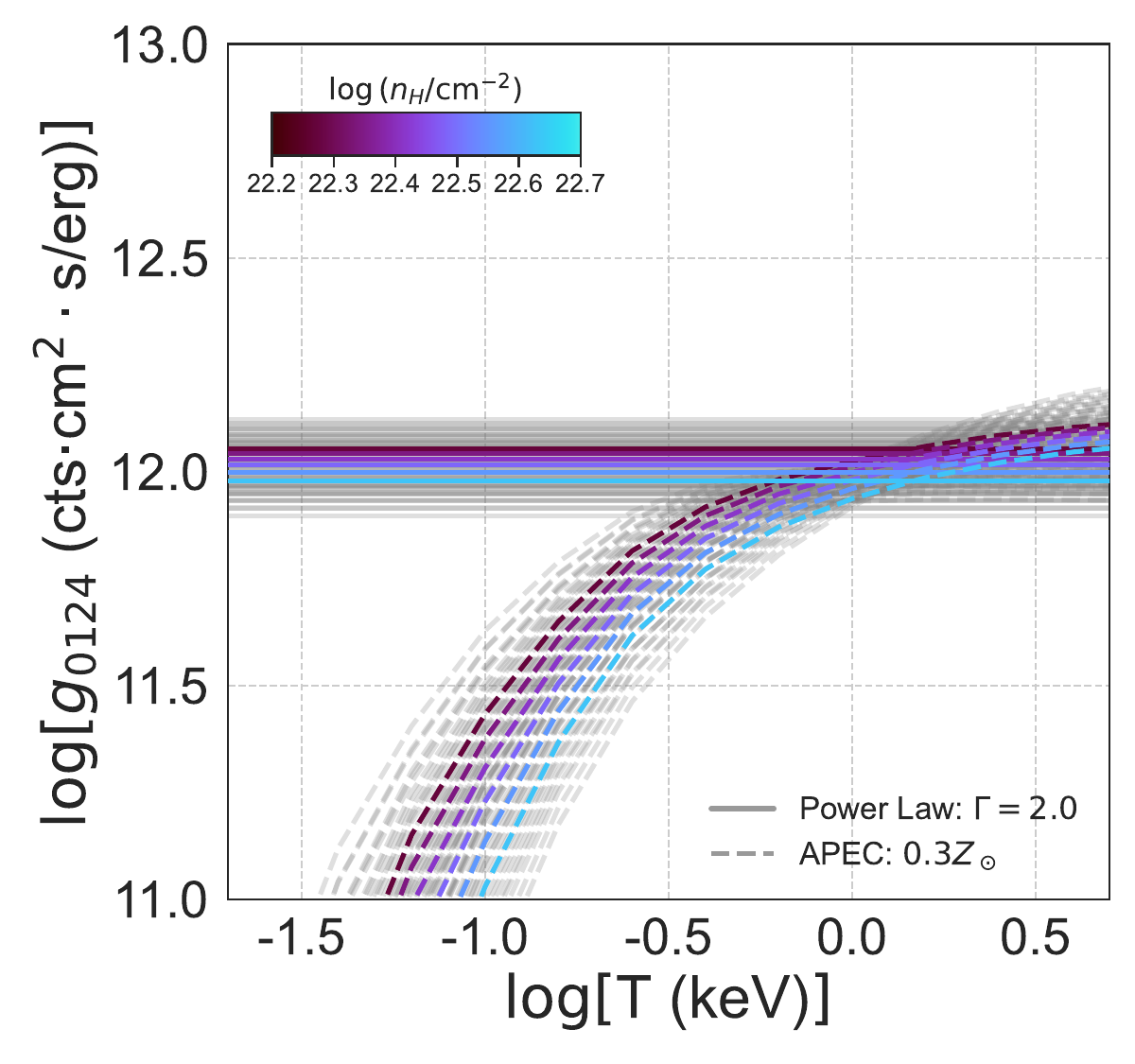}
  \caption{The ECFs that convert the rest-frame $0.1-2.4$ keV band flux with galactic absorption correction to the $0.2-2.3$ keV count rate as a function of gas temperature based on $Z = 0.3 Z_{\odot}$ APEC (dashed lines) and $\Gamma = 2.0$ power-law (solid lines) models with different parameters, respectively. Note that the galactic column density of neutral hydrogen are varied from  $\log \left(n_{\rm H}/{\rm cm}^{-2}\right) \simeq 20.26$ to $20.67$. We highlight the relations at redshift of $0.5$ color-coded by $\log \left(n_{\rm H}/{\rm cm}^{-2}\right)$. }
  \label{fig:ecf}
\end{figure}

In section~\ref{sec:xlums}, we calculate the ECFs based on power-law model of photon index $\Gamma = 2.0$ with a series of galactic absorptions and redshifts. The power-law model generally describe the X-ray emission of point sources such as AGNs \citep[e.g.,][]{Alexander..2011}. However, massive groups contain large amount of IGM on average. The IGM emission can be modeled with the thermal APEC model \citep{Smith..2001, Foster..2012}. To show which model can well describe the X-ray emission of our samples, we measure the hardness ratio (HR) from observed DESI groups and compare them with theoretical HRs based on different models, respectively. The HR is defined as follows:
\begin{equation}
{\rm HR} = \frac{N_{\rm src}^{\rm H} - N_{\rm src}^{\rm S}}{N_{\rm src}^{\rm H} + N_{\rm src}^{\rm S}},
\end{equation}
where $N_{\rm src}^{\rm H}$ and $N_{\rm src}^{\rm S}$ are the source counts after subtracting the background counts in the ranges $1.0-2.0$ keV and $0.2-1.0$ keV, respectively. 

To increase the significance of the results for observed groups, we derive the HRs using the stacks for groups with similar $M_h$ and $z_g$. The stacking methods for $N_{\rm src}^{\rm H}$ and $N_{\rm src}^{\rm S}$ are similar as used in stacking the images for faint groups (see section~\ref{sec:stack}), and the errors of stacked HR are propagated from their source count errors. 

In Figure~\ref{fig:hardness}, we plot the observed HR as a function of gas temperature. The gas temperature are derived based on $M_h - T$ relation given by \citet{Babyk..2023}. As can be seen, the HRs for groups with similar $M_h$ and $z_g$ are roughly lie in the range of $-0.4$ to $-0.1$. We also show the HRs given by various theoretical models taking into account the eFEDS effective area. Note that the results given by the power law models do not depend on gas temperature. We see that the theoretical HRs given by power law model with $\Gamma = 2.0$ can roughly represent the observed results in all of the $M_h$ range. We note that the results predicted by APEC and $\Gamma = 2.0$ power-law models show consistency when $\log \left[M_h/(h^{-1}M_\odot)\right] \gtrsim 13.5$. Moreover, the mechanics for X-ray emission of high- and low-mass systems might be different. Owing to the fact that the stacked results for massive groups are limited by small number statistics, we cannot clearly understand whether APEC or $\Gamma = 2.0$ power-law model can describe the X-ray of massive groups. However, as shown in figure~\ref{fig:ecf}, the difference between the ECFs obtained by both models are very small, at most $\lesssim 0.1$ dex at $T \gtrsim 3$ keV, thus we take use of the power-law model with $\Gamma = 2.0$ for simplicity.


\bsp	
\label{lastpage}

\end{document}